\documentclass[10pt,conference,letterpaper,nofonttune]{IEEEtran}
\IEEEoverridecommandlockouts
\usepackage{amsmath,amssymb,amsfonts}
\usepackage{amsthm}
\usepackage{multirow}

\usepackage{algorithmic}
\usepackage{graphicx}
\usepackage{textcomp}
\usepackage{xcolor}
\usepackage{xspace}
\usepackage{float}
\usepackage{enumitem} 
\usepackage{balance}
\usepackage{stfloats}
\usepackage{algorithm}
\usepackage{soul}
\usepackage{diagbox}

\allowdisplaybreaks  

\usepackage[font=footnotesize]{subcaption}
\usepackage[font=footnotesize]{caption}

\usepackage[acronyms,nonumberlist,nopostdot,nomain,nogroupskip,acronymlists={hidden}]{glossaries}
\newglossary[algh]{hidden}{acrh}{acnh}{Hidden Acronyms}
\usepackage[draft]{hyperref}
\usepackage[square, numbers, sort&compress]{natbib} 

\usepackage{tikz}
\usepackage{pgfplots}
\pgfplotsset{compat=newest}
\pgfplotsset{plot coordinates/math parser=false}
\newlength\fheight
\newlength\fwidth
\usetikzlibrary{plotmarks,patterns,decorations.pathreplacing,backgrounds,calc,arrows,arrows.meta,spy,matrix,scopes,pgfplots.colorbrewer}
\usepgfplotslibrary{patchplots,groupplots,colorbrewer}
\usepackage{tikzscale}

\newif\ifexttikz
\exttikzfalse

\ifexttikz
    \usetikzlibrary{external}
\fi

\def\name{ScalO-RAN\xspace}
\newcommand{\oran}{O-RAN\xspace}
\newcommand{\ran}{\gls{ran}\xspace}
\newcommand{\ai}{\gls{ai}\xspace}
\newcommand{\ml}{\gls{ml}\xspace}
\newcommand{\kpi}{\gls{kpi}\xspace}
\newcommand{\kpis}{\glspl{kpi}\xspace}
\newcommand{\nearrt}{\gls{near-rt}\xspace}
\newcommand{\nonrt}{\gls{non-rt}\xspace}
\newcommand{\reqs}{\mathcal{R}}
\newcommand{\apps}{\mathcal{A}}
\newcommand{\servers}{\mathcal{S}}
\newcommand{\ric}{\gls{ric}\xspace}
\newcommand{\rics}{\glspl{ric}\xspace}

\newtheorem{theorem}{Theorem}

\ifexttikz
\else
\usepackage{tikzpagenodes,etoolbox}
\usetikzlibrary{calc}
\usepackage[contents={}]{background}
\AddEverypageHook{%
\ifnumequal{\thepage}{1}{%
    \tikz[remember picture,overlay]{%
        \node[draw,
        minimum width=1.03\textwidth,
        text width=1.02\textwidth,
        font=\footnotesize
        ]
        at ($(current page header area) - (0,5pt)$)
        {%
        This paper has been accepted for publication on IEEE International Conference on Computer Communications (INFOCOM) 2024. This is the author's accepted version of the article. The final version published by IEEE is S. Maxenti, S. D'Oro, L. Bonati, M. Polese, A. Capone, and T. Melodia, ``\name: Energy-aware Network Intelligence Scaling in Open RAN'' \textit{IEEE INFOCOM 2024 - IEEE Conference on Computer Communications}, Vancouver, BC, Canada, 2024.
        };
        \node[draw,
        minimum width=1.03\textwidth,
        text width=1.02\textwidth,
        font=\footnotesize
        ]
        at (current page footer area)
        {%
        ©2024 IEEE. Personal use of this material is permitted. Permission from IEEE must be obtained for all other uses, in any current or future media, including reprinting/republishing this material for advertising or promotional purposes, creating new collective works, for resale or redistribution to servers or lists, or reuse of any copyrighted component of this work in other works.
        };
    }%
}{}
}
\fi

\begin{document}
\newacronym{rbac}{RBAC}{Role-based access control}
\newacronym{ci/cd}{CI/CD}{continuous integration and continuous deployment}
\newacronym{s2i}{S2I}{Source-To-Image}
\newacronym{sa}{SA}{Stand Alone}
\newacronym{sos}{SOS}{Special Order Set}
\newacronym[type=hidden]{open ran}{Open RAN}{Open \gls{ran}}
\newacronym{bts}{BTS}{Base Transceiver Station}
\newacronym{sc-fdma}{SC-FDMA}{Single Carrier-Frequency Division Multiple Access}
\newacronym{papr}{PAPR}{Peak-to-Average Power Ratio}
\newacronym{idrac}{iDRAC}{Integrated Dell Remote Access Controller}
\newacronym{wiot}{WIoT}{Institute for The Wireless Internet of Things}
\newacronym{saas}{SaaS}{Software-as-a-Service}
\newacronym{iaas}{IaaS}{Infrastructure-as-a-Service}
\newacronym{paas}{PaaS}{Platform-as-a-Service}
\newacronym{mrn}{MRN}{Mobile Radio Networks}
\newacronym{ran}{RAN}{Radio Access Network}
\newacronym{o-ran}{O-RAN}{}
\newacronym{cnn}{CNN}{Convolutional Neural Network}
\newacronym{sdn}{SDN}{Software Defined Networking}
\newacronym{nfv}{NFV}{Network Functions Virtualization}
\newacronym{appmgr}{APPMGR}{App Manager}
\newacronym{near-rt}{Near-RT}{Near-real-time}
\newacronym{non-rt}{Non-RT}{Non-real-time}
\newacronym{sla}{SLA}{Service Level Agreement}
\newacronym{3gpp}{3GPP}{3rd Generation Partnership Project}
\newacronym{cu}{CU}{Central Unit}
\newacronym{du}{DU}{Distributed Unit}
\newacronym{2g}{2G}{Second-Generation}
\newacronym{3g}{3G}{Third-Generation}
\newacronym{4g}{4G}{Fourth-generation}
\newacronym{5g}{5G}{Fifth-Generation}
\newacronym{6g}{6G}{Sixth-Generation}
\newacronym{5gc}{5GC}{5G Core}
\newacronym{adc}{ADC}{Analog to Digital Converter}
\newacronym{aerpaw}{AERPAW}{Aerial Experimentation and Research Platform for Advanced Wireless}
\newacronym{ai}{AI}{Artificial Intelligence}
\newacronym{aimd}{AIMD}{Additive Increase Multiplicative Decrease}
\newacronym{am}{AM}{Acknowledged Mode}
\newacronym{amc}{AMC}{Adaptive Modulation and Coding}
\newacronym{ilp}{ILP}{Integer Linear Programming}
\newacronym{milp}{MILP}{Mixed Integer Linear Programming}
\newacronym{amf}{AMF}{Access and Mobility Management Function}
\newacronym{aops}{AOPS}{Adaptive Order Prediction Scheduling}
\newacronym{api}{API}{Application Programming Interface}
\newacronym{apn}{APN}{Access Point Name}
\newacronym{ap}{AP}{Application Protocol}
\newacronym{aqm}{AQM}{Active Queue Management}
\newacronym{asn1}{ASN.1}{Abstract Syntax Notation One}
\newacronym{ausf}{AUSF}{Authentication Server Function}
\newacronym{avc}{AVC}{Advanced Video Coding}
\newacronym{awgn}{AGWN}{Additive White Gaussian Noise}
\newacronym{balia}{BALIA}{Balanced Link Adaptation Algorithm}
\newacronym{bbu}{BBU}{Base Band Unit}
\newacronym{bdp}{BDP}{Bandwidth-Delay Product}
\newacronym{ber}{BER}{Bit Error Rate}
\newacronym{bf}{BF}{Beamforming}
\newacronym{bler}{BLER}{Block Error Rate}
\newacronym{brr}{BRR}{Bayesian Ridge Regressor}
\newacronym{bs}{BS}{Base Station}
\newacronym{bsr}{BSR}{Buffer Status Report}
\newacronym{bss}{BSS}{Business Support System}
\newacronym{ca}{CA}{Carrier Aggregation}
\newacronym{caas}{CaaS}{Connectivity-as-a-Service}
\newacronym{cb}{CB}{Code Block}
\newacronym{cc}{CC}{Congestion Control}
\newacronym{compc}{CC}{Component Carrier}
\newacronym{ccid}{CCID}{Congestion Control ID}
\newacronym{cco}{CC}{Carrier Component}
\newacronym{cdd}{CDD}{Cyclic Delay Diversity}
\newacronym{cdf}{CDF}{Cumulative Distribution Function}
\newacronym{cdn}{CDN}{Content Distribution Network}
\newacronym{cn}{CN}{Core Network}
\newacronym{codel}{CoDel}{Controlled Delay Management}
\newacronym{comac}{COMAC}{Converged Multi-Access and Core}
\newacronym{cord}{CORD}{Central Office Re-architected as a Datacenter}
\newacronym{cornet}{CORNET}{COgnitive Radio NETwork}
\newacronym{cosmos}{COSMOS}{Cloud Enhanced Open Software Defined Mobile Wireless Testbed for City-Scale Deployment}
\newacronym{cots}{COTS}{Commercial Off-the-Shelf}
\newacronym{cp}{CP}{Control Plane}
\newacronym{cpu}{CPU}{Central Processing Unit}
\newacronym{cqi}{CQI}{Channel Quality Information}
\newacronym{cr}{CR}{Cognitive Radio}
\newacronym{cql}{CQL}{Conservative Q-Learning}
\newacronym{cran}{CRAN}{Cloud \gls{ran}}
\newacronym{crs}{CRS}{Cell Reference Signal}
\newacronym{csi}{CSI}{Channel State Information}
\newacronym{csirs}{CSI-RS}{Channel State Information - Reference Signal}
\newacronym{cucp}{CU-CP}{Centralized Unit - Control Plane}
\newacronym{cuup}{CU-UP}{Centralized Unit - User Plane}
\newacronym{d2tcp}{D$^2$TCP}{Deadline-aware Data center TCP}
\newacronym{d2d}{D2D}{Device to Device}
\newacronym{d3}{D$^3$}{Deadline-Driven Delivery} 
\newacronym{dac}{DAC}{Digital to Analog Converter}
\newacronym{dag}{DAG}{Directed Acyclic Graph}
\newacronym{das}{DAS}{Distributed Antenna System}
\newacronym{dash}{DASH}{Dynamic Adaptive Streaming over HTTP}
\newacronym{dc}{DC}{Dual Connectivity}
\newacronym{dccp}{DCCP}{Datagram Congestion Control Protocol}
\newacronym{dce}{DCE}{Direct Code Execution}
\newacronym{dci}{DCI}{Downlink Control Information}
\newacronym{dctcp}{DCTCP}{Data Center TCP}
\newacronym{dl}{DL}{Downlink}
\newacronym{dmr}{DMR}{Deadline Miss Ratio}
\newacronym{dmrs}{DMRS}{DeModulation Reference Signal}
\newacronym{drl}{DRL}{Deep Reinforcement Learning}
\newacronym{drlcc}{DRL-CC}{Deep Reinforcement Learning Congestion Control}
\newacronym{drb}{DRB}{Data Radio Bearer}
\newacronym{drs}{DRS}{Discovery Reference Signal}
\newacronym{dnn}{DNN}{Deep Neural Network}
\newacronym{dqn}{DQN}{Deep Q-Network}
\newacronym{e2e}{E2E}{end-to-end}
\newacronym{e2ap}{E2AP}{E2 Application Protocol}
\newacronym{e2sm}{E2SM}{E2 Service Model}
\newacronym{ecaas}{ECaaS}{Edge-Cloud-as-a-Service}
\newacronym{ecn}{ECN}{Explicit Congestion Notification}
\newacronym{edc}{EDC}{Edge Data Center}
\newacronym{edf}{EDF}{Earliest Deadline First}
\newacronym{embb}{eMBB}{Enhanced Mobile Broadband}
\newacronym{empower}{EMPOWER}{EMpowering transatlantic PlatfOrms for advanced WirEless Research}
\newacronym{enb}{eNB}{evolved Node Base}
\newacronym{endc}{EN-DC}{E-UTRAN-NR Dual Connectivity}
\newacronym{epc}{EPC}{Evolved Packet Core}
\newacronym{eps}{EPS}{Evolved Packet System}
\newacronym{es}{ES}{Edge Server}
\newacronym{etl}{ETL}{Extract, Transform and Load}
\newacronym{etsi}{ETSI}{European Telecommunications Standards Institute}
\newacronym[firstplural=Estimated Times of Arrival (ETAs)]{eta}{ETA}{Estimated Time of Arrival}
\newacronym{eutran}{E-UTRAN}{Evolved Universal Terrestrial Access Network}
\newacronym{faas}{FaaS}{Function-as-a-Service}
\newacronym{fapi}{FAPI}{Functional Application Platform Interface}
\newacronym{fcaps}{FCAPS}{Fault, Configuration, Accounting, Performance and Security}
\newacronym{fdd}{FDD}{Frequency Division Duplexing}
\newacronym{fdm}{FDM}{Frequency Division Multiplexing}
\newacronym{fdma}{FDMA}{Frequency Division Multiple Access}
\newacronym{fed4fire}{FED4FIRE+}{Federation 4 Future Internet Research and Experimentation Plus}
\newacronym{fir}{FIR}{Finite Impulse Response}
\newacronym{fit}{FIT}{Future \acrlong{iot}}
\newacronym{fpga}{FPGA}{Field Programmable Gate Array}
\newacronym{fr1}{FR1}{Frequency Range 1}
\newacronym{fr2}{FR2}{Frequency Range 2}
\newacronym{fs}{FS}{Fast Switching}
\newacronym{fscc}{FSCC}{Flow Sharing Congestion Control}
\newacronym{ftp}{FTP}{File Transfer Protocol}
\newacronym{fw}{FW}{Flow Window}
\newacronym{gbr}{GBR}{Guaranteed Bit Rate}
\newacronym{ge}{GE}{Gaussian Elimination}
\newacronym{gnb}{gNB}{Next Generation Node Base}
\newacronym{gop}{GOP}{Group of Pictures}
\newacronym{gpr}{GPR}{Gaussian Process Regressor}
\newacronym{gpu}{GPU}{Graphics Processing Unit}
\newacronym{gtp}{GTP}{GPRS Tunneling Protocol}
\newacronym{gprs}{GPRS}{General Packet Radio Service}
\newacronym{gtpc}{GTP-C}{GPRS Tunnelling Protocol Control Plane}
\newacronym{gtpu}{GTP-U}{GPRS Tunnelling Protocol User Plane}
\newacronym{gtpv2c}{GTPv2-C}{\gls{gtp} v2 - Control}
\newacronym{gw}{GW}{Gateway}
\newacronym{harq}{HARQ}{Hybrid Automatic Repeat reQuest}
\newacronym{hetnet}{HetNet}{Heterogeneous Network}
\newacronym{hh}{HH}{Hard Handover}
\newacronym{ho}{HO}{Handover}
\newacronym{hol}{HOL}{Head-of-Line}
\newacronym{hsdpa}{HSDPA}{High-Speed Downlink Packet Access}
\newacronym{hsupa}{HSUPA}{High-Speed Uplink Packet Access}
\newacronym{hqf}{HQF}{Highest-quality-first}
\newacronym{hss}{HSS}{Home Subscription Server}
\newacronym{http}{HTTP}{HyperText Transfer Protocol}
\newacronym{ia}{IA}{Initial Access}
\newacronym{iab}{IAB}{Integrated Access and Backhaul}
\newacronym{ic}{IC}{Incident Command}
\newacronym{ietf}{IETF}{Internet Engineering Task Force}
\newacronym{imsi}{IMSI}{International Mobile Subscriber Identity}
\newacronym{imt}{IMT}{International Mobile Telecommunication}
\newacronym{ims}{IMS}{\gls{ip} Multimedia Settings}
\newacronym{iot}{IoT}{Internet of Things}
\newacronym{ip}{IP}{Internet Protocol}
\newacronym{itu}{ITU}{International Telecommunication Union}
\newacronym{kpi}{KPI}{Key Performance Indicator}
\newacronym{kpm}{KPM}{Key Performance Measurement}
\newacronym{ni}{NI}{Network Interfaces}
\newacronym{kvm}{KVM}{Kernel-based Virtual Machine}
\newacronym{los}{LOS}{Line-of-Sight}
\newacronym{ldc}{LDC}{Local Data Center}
\newacronym{lsm}{LSM}{Link-to-System Mapping}
\newacronym{lstm}{LSTM}{Long Short Term Memory}
\newacronym{lte}{LTE}{Long Term Evolution}
\newacronym{lxc}{LXC}{Linux Container}
\newacronym{m2m}{M2M}{Machine to Machine}
\newacronym{mac}{MAC}{Medium Access Control}
\newacronym{manet}{MANET}{Mobile Ad Hoc Network}
\newacronym{mano}{MANO}{Management and Orchestration}
\newacronym{mbr}{MBR}{Maximum Bit Rate}
\newacronym{mc}{MC}{Multi-Connectivity}
\newacronym{mcc}{MCC}{Mobile Cloud Computing}
\newacronym{mchem}{MCHEM}{Massive Channel Emulator}
\newacronym{mcs}{MCS}{Modulation and Coding Scheme}
\newacronym{mdp}{MDP}{Markov Decision Process}
\newacronym{mec}{MEC}{Multi-access Edge Computing}
\newacronym{mec2}{MEC}{Mobile Edge Cloud}
\newacronym{mfc}{MFC}{Mobile Fog Computing}
\newacronym{mgen}{MGEN}{Multi-Generator}
\newacronym{mi}{MI}{Mutual Information}
\newacronym{mib}{MIB}{Master Information Block}
\newacronym{minlp}{MINLP}{Mixed Integer Non-Linear Problem}
\newacronym{miesm}{MIESM}{Mutual Information Based Effective SINR}
\newacronym{mimo}{MIMO}{Multiple Input, Multiple Output}
\newacronym{ml}{ML}{Machine Learning}
\newacronym{mlp}{MLP}{Mixed Linear Programming}
\newacronym{mlr}{MLR}{Maximum-local-rate}
\newacronym[plural=\gls{mme}s,firstplural=Mobility Management Entities (MMEs)]{mme}{MME}{Mobility Management Entity}
\newacronym{mmtc}{mMTC}{Massive Machine-Type Communications}
\newacronym{mmwave}{mmWave}{millimeter wave}
\newacronym{mpdccp}{MP-DCCP}{Multipath Datagram Congestion Control Protocol}
\newacronym{mptcp}{MPTCP}{Multipath TCP}
\newacronym{mr}{MR}{Maximum Rate}
\newacronym{mrdc}{MR-DC}{Multi \gls{rat} \gls{dc}}
\newacronym{mse}{MSE}{Mean Square Error}
\newacronym{mss}{MSS}{Maximum Segment Size}
\newacronym{mt}{MT}{Mobile Termination}
\newacronym{mtd}{MTD}{Machine-Type Device}
\newacronym{mtu}{MTU}{Maximum Transmission Unit}
\newacronym{mumimo}{MU-MIMO}{Multi-user \gls{mimo}}
\newacronym{mvno}{MVNO}{Mobile Virtual Network Operator}
\newacronym{nalu}{NALU}{Network Abstraction Layer Unit}
\newacronym{nas}{NAS}{Non-Access Stratum}
\newacronym{ngran}{NG-RAN}{Next Generation - \gls{ran}}
\newacronym{ns3}{ns-3}{Network Simulator 3}
\newacronym{nbiot}{NB-IoT}{Narrow Band IoT}
\newacronym{nf}{NF}{Network Function}
\newacronym{nfvi}{NFVI}{Network Function Virtualization Infrastructure}
\newacronym{nic}{NIC}{Network Interface Card}
\newacronym{nlos}{NLOS}{Non-Line-of-Sight}
\newacronym{now}{NOW}{Non Overlapping Window}
\newacronym{nsm}{NSM}{Network Service Mesh}
\newacronym{nrf}{NRF}{Network Repository Function}
\newacronym{nsa}{NSA}{Non Stand Alone}
\newacronym{nse}{NSE}{Network Slicing Engine}
\newacronym{nssf}{NSSF}{Network Slice Selection Function}
\newacronym{o2i}{O2I}{Outdoor to Indoor}
\newacronym{oai}{OAI}{OpenAirInterface}
\newacronym{oaicn}{OAI-CN}{\gls{oai} \acrlong{cn}}
\newacronym{oairan}{OAI-RAN}{\acrlong{oai} \acrlong{ran}}
\newacronym{oam}{OAM}{Operations, Administration and Maintenance}
\newacronym{ofdm}{OFDM}{Orthogonal Frequency Division Multiplexing}
\newacronym{ofdma}{OFDMA}{Orthogonal Frequency Division Multiple Access}
\newacronym{olia}{OLIA}{Opportunistic Linked Increase Algorithm}
\newacronym{omec}{OMEC}{Open Mobile Evolved Core}
\newacronym{onap}{ONAP}{Open Network Automation Platform}
\newacronym{onf}{ONF}{Open Networking Foundation}
\newacronym{onos}{ONOS}{Open Networking Operating System}
\newacronym{oom}{OOM}{\gls{onap} Operations Manager}
\newacronym{opnfv}{OPNFV}{Open Platform for \gls{nfv}}
\newacronym{orbit}{ORBIT}{Open-Access Research Testbed for Next-Generation Wireless Networks}
\newacronym{os}{OS}{Operating System}
\newacronym{oss}{OSS}{Operations Support System}
\newacronym{pa}{PA}{Position-aware}
\newacronym{pase}{PASE}{Prioritization, Arbitration, and Self-adjusting Endpoints}
\newacronym{pawr}{PAWR}{Platforms for Advanced Wireless Research}
\newacronym{pbch}{PBCH}{Physical Broadcast Channel}
\newacronym{pcell}{PCell}{Primary Cell}
\newacronym{pcef}{PCEF}{Policy and Charging Enforcement Function}
\newacronym{pcfich}{PCFICH}{Physical Control Format Indicator Channel}
\newacronym{pcrf}{PCRF}{Policy and Charging Rules Function}
\newacronym{pdcch}{PDCCH}{Physical Downlink Control Channel}
\newacronym{pdcp}{PDCP}{Packet Data Convergence Protocol}
\newacronym{pdcpc}{PDCP-C}{Packet Data Convergence Protocol - Control Plane}
\newacronym{pdcpu}{PDCP-U}{Packet Data Convergence Protocol - User Plane}
\newacronym{pdsch}{PDSCH}{Physical Downlink Shared Channel}
\newacronym{pdu}{PDU}{Packet Data Unit}
\newacronym{pf}{PF}{Proportional Fair}
\newacronym{pgw}{PGW}{Packet Gateway}
\newacronym{phich}{PHICH}{Physical Hybrid ARQ Indicator Channel}
\newacronym{phy}{PHY}{Physical}
\newacronym{phyu}{PHY-U}{Upper Physical}
\newacronym{phyl}{PHY-L}{Lower Physical}
\newacronym{pmch}{PMCH}{Physical Multicast Channel}
\newacronym{pmi}{PMI}{Precoding Matrix Indicators}
\newacronym{powder}{POWDER}{Platform for Open Wireless Data-driven Experimental Research}
\newacronym{ppo}{PPO}{Proximal Policy Optimization}
\newacronym{ppp}{PPP}{Poisson Point Process}
\newacronym{prach}{PRACH}{Physical Random Access Channel}
\newacronym{prb}{PRB}{Physical Resource Block}
\newacronym{pscell}{PSCell}{Primary cell of the Secondary Node}
\newacronym{psnr}{PSNR}{Peak Signal to Noise Ratio}
\newacronym{pss}{PSS}{Primary Synchronization Signal}
\newacronym{pucch}{PUCCH}{Physical Uplink Control Channel}
\newacronym{pusch}{PUSCH}{Physical Uplink Shared Channel}
\newacronym{qam}{QAM}{Quadrature Amplitude Modulation}
\newacronym{qci}{QCI}{\gls{qos} Class Identifier}
\newacronym{qcqp}{QCQP}{Quadratically Constrained Quadratic Problem}
\newacronym{qoe}{QoE}{Quality of Experience}
\newacronym{qos}{QoS}{Quality of Service}
\newacronym{quic}{QUIC}{Quick UDP Internet Connections}
\newacronym{rach}{RACH}{Random Access Channel}
\newacronym[firstplural=Radio Access Technologies (RATs)]{rat}{RAT}{Radio Access Technology}
\newacronym{rest}{REST}{REpresentational State Transfer}
\newacronym{rcn}{RCN}{Research Coordination Network}
\newacronym{rec}{REC}{Radio Edge Cloud}
\newacronym{red}{RED}{Random Early Detection}
\newacronym{rem}{REM}{Random Ensemble Mixture}
\newacronym{renew}{RENEW}{Reconfigurable Eco-system for Next-generation End-to-end Wireless}
\newacronym{rf}{RF}{Radio Frequency}
\newacronym{rfc}{RFC}{Request for Comments}
\newacronym{rfr}{RFR}{Random Forest Regressor}
\newacronym{ric}{RIC}{RAN Intelligent Controller}
\newacronym{rlc}{RLC}{Radio Link Control}
\newacronym{rl}{RL}{Reinforcement Learning}
\newacronym{rlf}{RLF}{Radio Link Failure}
\newacronym{rlnc}{RLNC}{Random Linear Network Coding}
\newacronym{rmr}{RMR}{RIC Message Routing}
\newacronym{rmse}{RMSE}{Root Mean Squared Error}
\newacronym{rnis}{RNIS}{Radio Network Information Service}
\newacronym{rnib}{RNIB}{Radio Network Information Base}
\newacronym{rr}{RR}{Round Robin}
\newacronym{rrc}{RRC}{Radio Resource Control}
\newacronym{rrm}{RRM}{Radio Resource Management}
\newacronym{rru}{RRU}{Remote Radio Unit}
\newacronym{rs}{RS}{Remote Server}
\newacronym{rsrp}{RSRP}{Reference Signal Received Power}
\newacronym{rsrq}{RSRQ}{Reference Signal Received Quality}
\newacronym{rss}{RSS}{Received Signal Strength}
\newacronym{rssi}{RSSI}{Received Signal Strength Indicator}
\newacronym{rtt}{RTT}{Round Trip Time}
\newacronym{rt}{RT}{Real-time}
\newacronym{ru}{RU}{Radio Unit}
\newacronym{rw}{RW}{Receive Window}
\newacronym{rx}{RX}{Receiver}
\newacronym{s1ap}{S1AP}{S1 Application Protocol}
\newacronym{sack}{SACK}{Selective Acknowledgment}
\newacronym{sap}{SAP}{Service Access Point}
\newacronym{sc2}{SC2}{Spectrum Collaboration Challenge}
\newacronym{scef}{SCEF}{Service Capability Exposure Function}
\newacronym{sch}{SCH}{Secondary Cell Handover}
\newacronym{scoot}{SCOOT}{Split Cycle Offset Optimization Technique}
\newacronym{sctp}{SCTP}{Stream Control Transmission Protocol}
\newacronym{sdap}{SDAP}{Service Data Adaptation Protocol}
\newacronym{sdk}{SDK}{Software Development Kit}
\newacronym{sdm}{SDM}{Space Division Multiplexing}
\newacronym{sdma}{SDMA}{Spatial Division Multiple Access}
\newacronym{sdr}{SDR}{Software-defined Radio}
\newacronym{seba}{SEBA}{SDN-Enabled Broadband Access}
\newacronym{sgsn}{SGSN}{Serving GPRS Support Node}
\newacronym{sgw}{SGW}{Serving Gateway}
\newacronym{si}{SI}{Study Item}
\newacronym{sib}{SIB}{Secondary Information Block}
\newacronym{sinr}{SINR}{Signal to Interference plus Noise Ratio}
\newacronym{sip}{SIP}{Session Initiation Protocol}
\newacronym{siso}{SISO}{Single Input, Single Output}
\newacronym{sm}{SM}{Service Model}
\newacronym{smf}{SMF}{Session Management Function}
\newacronym{smo}{SMO}{Service Management and Orchestration}
\newacronym{sms}{SMS}{Short Message Service}
\newacronym{smsgmsc}{SMS-GMSC}{\gls{sms}-Gateway}
\newacronym{snr}{SNR}{Signal-to-Noise-Ratio}
\newacronym{son}{SON}{Self-Organizing Network}
\newacronym{sptcp}{SPTCP}{Single Path TCP}
\newacronym{srb}{SRB}{Service Radio Bearer}
\newacronym{srn}{SRN}{Standard Radio Node}
\newacronym{srs}{SRS}{Sounding Reference Signal}
\newacronym{ss}{SS}{Synchronization Signal}
\newacronym{sss}{SSS}{Secondary Synchronization Signal}
\newacronym{st}{ST}{Spanning Tree}
\newacronym{svc}{SVC}{Scalable Video Coding}
\newacronym{tb}{TB}{Transport Block}
\newacronym{tcp}{TCP}{Transmission Control Protocol}
\newacronym{tdd}{TDD}{Time Division Duplexing}
\newacronym{tdm}{TDM}{Time Division Multiplexing}
\newacronym{tdma}{TDMA}{Time Division Multiple Access}
\newacronym{cdma}{CDMA}{Code Division Multiple Access}
\newacronym{tfl}{TfL}{Transport for London}
\newacronym{tfrc}{TFRC}{TCP-Friendly Rate Control}
\newacronym{tft}{TFT}{Traffic Flow Template}
\newacronym{tgen}{TGEN}{Traffic Generator}
\newacronym{tip}{TIP}{Telecom Infra Project}
\newacronym{tm}{TM}{Transparent Mode}
\newacronym{tco}{TCO}{total cost of ownership}
\newacronym{to}{TO}{Telco Operator}
\newacronym{tr}{TR}{Technical Report}
\newacronym{trp}{TRP}{Transmitter Receiver Pair}
\newacronym{ts}{TS}{Traffic Steering}
\newacronym{tti}{TTI}{Transmission Time Interval}
\newacronym{ttt}{TTT}{Time-to-Trigger}
\newacronym{tx}{TX}{Transmitter}
\newacronym{uas}{UAS}{Unmanned Aerial System}
\newacronym{uav}{UAV}{Unmanned Aerial Vehicle}
\newacronym{udm}{UDM}{Unified Data Management}
\newacronym{udp}{UDP}{User Datagram Protocol}
\newacronym{udr}{UDR}{Unified Data Repository}
\newacronym{ue}{UE}{User Equipment}
\newacronym{uhd}{UHD}{\gls{usrp} Hardware Driver}
\newacronym{ul}{UL}{Uplink}
\newacronym{um}{UM}{Unacknowledged Mode}
\newacronym{uml}{UML}{Unified Modeling Language}
\newacronym{upa}{UPA}{Uniform Planar Array}
\newacronym{upf}{UPF}{User Plane Function}
\newacronym{urllc}{URLLC}{Ultra Reliable and Low Latency Communications}
\newacronym{usa}{U.S.}{United States}
\newacronym{usim}{USIM}{Universal Subscriber Identity Module}
\newacronym{usrp}{USRP}{Universal Software Radio Peripheral}
\newacronym{utc}{UTC}{Urban Traffic Control}
\newacronym{vim}{VIM}{Virtualization Infrastructure Manager}
\newacronym{vm}{VM}{Virtual Machine}
\newacronym{vnf}{VNF}{virtual network function}
\newacronym{nr}{NR}{New Radio}
\newacronym{volte}{VoLTE}{Voice over LTE}
\newacronym{vonr}{VoNR}{Voice over NR}
\newacronym{voltha}{VOLTHA}{Virtual OLT HArdware Abstraction}
\newacronym{vr}{VR}{Virtual Reality}
\newacronym{vran}{vRAN}{Virtualized \gls{ran}}
\newacronym{vss}{VSS}{Video Streaming Server}
\newacronym{v2x}{V2X}{Vehicle-to-everything}
\newacronym{wbf}{WBF}{Wired Bias Function}
\newacronym{wf}{WF}{Waterfilling}
\newacronym{wlan}{WLAN}{Wireless Local Area Network}
\newacronym{osm}{OSM}{Open Source \gls{nfv} Management and Orchestration}
\newacronym{pnf}{PNF}{Physical Network Function}
\newacronym{mtc}{MTC}{Machine-type Communications}
\newacronym{osc}{OSC}{O-RAN Software Community}
\newacronym{rc}{RC}{RAN Control}
\newacronym{ar}{AR}{Augmented Reality}
\newacronym{daps}{DAPS}{Dual Active Protocol Stack}
\newacronym{nib}{NIB}{Network Information Base}
\newacronym{isa}{ISA}{Instruction Set Architecture}
\newacronym{abi}{ABI}{Application Binary Interface}
\newacronym{vbs}{vBS}{Virtual Base Station}
\newacronym{up}{UP}{User Plane}
\newacronym{ci}{CI}{Continuous Integration}
\newacronym{cd}{CD}{Continuous Deployment}
\tikzstyle{startstop} = [rectangle, rounded corners, minimum width=2cm, minimum height=0.5cm,text centered, draw=black]
\tikzstyle{io} = [trapezium, trapezium left angle=70, trapezium right angle=110, minimum width=3cm, minimum height=1cm, text centered, draw=black]
\tikzstyle{process} = [rectangle, minimum width=2cm, minimum height=0.5cm, text centered, draw=black, alignb=center]
\tikzstyle{decision} = [ellipse, minimum width=2cm, minimum height=1cm, text centered, draw=black]
\tikzstyle{arrow} = [thick,<->,>=stealth]
\tikzstyle{line} = [thick,>=stealth]
\tikzstyle{darrow} = [thick,<->,>=stealth,dashed]
\tikzstyle{sarrow} = [thick,->,>=stealth]
\tikzstyle{larrow} = [line width=0.1mm,dashdotted,->,>=stealth]
\tikzstyle{llarrow} = [line width=0.1mm,->,>=stealth]

\makeatletter
\def\grd@save@target#1{%
  \def\grd@target{#1}}
\def\grd@save@start#1{%
  \def\grd@start{#1}}
\tikzset{
  grid with coordinates/.style={
    to path={%
      \pgfextra{%
        \edef\grd@@target{(\tikztotarget)}%
        \tikz@scan@one@point\grd@save@target\grd@@target\relax
        \edef\grd@@start{(\tikztostart)}%
        \tikz@scan@one@point\grd@save@start\grd@@start\relax
        \draw[minor help lines] (\tikztostart) grid (\tikztotarget);
        \draw[major help lines] (\tikztostart) grid (\tikztotarget);
        \grd@start
        \pgfmathsetmacro{\grd@xa}{\the\pgf@x/1cm}
        \pgfmathsetmacro{\grd@ya}{\the\pgf@y/1cm}
        \grd@target
        \pgfmathsetmacro{\grd@xb}{\the\pgf@x/1cm}
        \pgfmathsetmacro{\grd@yb}{\the\pgf@y/1cm}
        \pgfmathsetmacro{\grd@xc}{\grd@xa + \pgfkeysvalueof{/tikz/grid with coordinates/major step x}}
        \pgfmathsetmacro{\grd@yc}{\grd@ya + \pgfkeysvalueof{/tikz/grid with coordinates/major step y}}
        \foreach \x in {\grd@xa,\grd@xc,...,\grd@xb}
        \node[anchor=north] at (\x,\grd@ya) {\pgfmathprintnumber{\x}};
        \foreach \y in {\grd@ya,\grd@yc,...,\grd@yb}
        \node[anchor=east] at (\grd@xa,\y) {\pgfmathprintnumber{\y}};
      }
    }
  },
  minor help lines/.style={
    help lines,
    gray,
    line cap =round,
    xstep=\pgfkeysvalueof{/tikz/grid with coordinates/minor step x},
    ystep=\pgfkeysvalueof{/tikz/grid with coordinates/minor step y}
  },
  major help lines/.style={
    help lines,
    line cap =round,
    line width=\pgfkeysvalueof{/tikz/grid with coordinates/major line width},
    xstep=\pgfkeysvalueof{/tikz/grid with coordinates/major step x},
    ystep=\pgfkeysvalueof{/tikz/grid with coordinates/major step y}
  },
  grid with coordinates/.cd,
  minor step x/.initial=.5,
  minor step y/.initial=.2,
  major step x/.initial=1,
  major step y/.initial=1,
  major line width/.initial=1pt,
}
\makeatother

\definecolor{desireRed}{RGB}{230,57,60}%
\definecolor{darkPurple}{RGB}{59,31,43}%
\definecolor{springGreen}{RGB}{37,223,145}%
\definecolor{queenBlue}{RGB}{69,123,157}%
\definecolor{spaceCadet}{RGB}{29,53,87}%

\def\si{\tikz\fill[scale=0.4](0,.35) -- (.25,0) -- (1,.7) -- (.25,.15) -- cycle;}

\title{\name: Energy-aware\\ Network Intelligence Scaling in Open RAN}

\author{\IEEEauthorblockN{
Stefano Maxenti\IEEEauthorrefmark{1}\IEEEauthorrefmark{2},
Salvatore D'Oro\IEEEauthorrefmark{1},
Leonardo Bonati\IEEEauthorrefmark{1},
Michele Polese\IEEEauthorrefmark{1},
Antonio Capone\IEEEauthorrefmark{2},
Tommaso Melodia\IEEEauthorrefmark{1}}
\IEEEauthorblockA{
\IEEEauthorrefmark{1}Institute for the Wireless Internet of Things, Northeastern University, Boston, MA, U.S.A.\\
\IEEEauthorrefmark{2}Politecnico di Milano, Milan, Italy\\
Email: \{maxenti.s, s.doro, l.bonati, m.polese, melodia\}@northeastern.edu, antonio.capone@polimi.it}
%
\thanks{This work was partially supported by the National Telecommunications and Information Administration (NTIA)'s Public Wireless Supply Chain Innovation Fund (PWSCIF) under Award No. 25-60-IF002,
and by OUSD(R\&E) through Army Research Laboratory Cooperative Agreement Number W911NF-19-2-0221. The views and conclusions contained in this document are those of the authors and should not be interpreted as representing the official policies, either expressed or implied, of the Army Research Laboratory or the U.S. Government. The U.S. Government is authorized to reproduce and distribute reprints for Government purposes notwithstanding any copyright notation herein.}
}

\setlist[itemize]{leftmargin=5.5mm}


\makeatletter
\patchcmd{\@maketitle}
{\addvspace{0.5\baselineskip}\egroup}
{\addvspace{-1.5\baselineskip}\egroup}
{}
{}
\makeatother
\maketitle
\begin{abstract}
%
Network virtualization, software-defined infrastructure, and orchestration are pivotal elements in contemporary networks, yielding new vectors for optimization and novel capabilities. In line with these principles, O-RAN presents an avenue to bypass vendor lock-in, circumvent vertical configurations, enable network programmability, and facilitate integrated artificial intelligence (AI) support.
Moreover, modern container orchestration frameworks (e.g., Kubernetes, Red Hat OpenShift) simplify the way cellular base stations, as well as the newly introduced \glspl{ric}, are deployed, managed, and orchestrated.
While this enables cost reduction via infrastructure sharing, it also makes it more challenging to meet O-RAN control latency requirements, especially during peak resource utilization. For instance, the Near-real-time \gls{ric} is in charge of executing applications (xApps) that must take control decisions within one second, and we show that container platforms available today fail in guaranteeing such timing constraints.
To address this problem, we propose \name, a control framework rooted in optimization and designed as an O-RAN rApp that allocates and scales AI-based O-RAN applications (xApps, rApps, dApps) to: (i) abide by application-specific latency requirements, and (ii) monetize the shared infrastructure while reducing energy consumption. 
%
We prototype \name on an OpenShift cluster with base stations,
\gls{ric}, and a set of AI-based xApps deployed as micro-services. We evaluate \name both numerically and experimentally. Our results show that \name can optimally allocate and distribute O-RAN applications within available computing nodes to accommodate even stringent latency requirements.
More importantly, we show that scaling O-RAN applications is primarily a time-constrained problem rather than a resource-constrained one, where scaling policies must account for stringent inference time of AI applications, and not only how many resources they consume.
%
%


\end{abstract}


\glsresetall


\section{Introduction}
\label{section:introduction}%


The need for more flexible, energy-efficient, and cost-effective cellular networks---capable at the same time of delivering and guaranteeing high data rates and low latency---is driving the telco ecosystem toward \ran cloudification. 
The shift leverages the principles of virtualization and softwarization, concepts deeply ingrained in cloud-computing and internetworking fields via \gls{sdn} and \gls{nfv}. These principles enable the design, development, and deployment of cellular networks with superior flexibility, which can be effectively monitored, controlled, optimized, upgraded, and reconfigured in real time via software.

This ongoing industry transformation has led to the Open \ran paradigm, and the creation of the \oran Alliance~\cite{polese2022understanding}. \oran leverages the principles described above to foster a cloud-based cellular architecture, with interoperable multi-vendor hardware and software components interconnected via open and standardized interfaces.
%
It also embeds \ai and \ml directly into the network to \textit{forecast} loads, \kpis and user mobility, \textit{control} \ran functionalities and spectrum usage, and \textit{classify} traffic profiles and identify anomalies, to name a few~\cite{brik2022deep,polese2022understanding}.
To enable flexible \acrshort{5g}/\acrshort{6g} networks, \oran introduces the concept of \gls{ric}, i.e., an abstraction enabling the execution of third-party network functions for \ai-based inference and control.
%
%
\glspl{ric} are based on micro-services embedding intelligent workloads, called xApps and rApps. \oran defines specifically the \nearrt (hosting xApps) and the \nonrt \glspl{ric} (hosting rApps) for inference loops up to $1$\:s and beyond $1$\:s, respectively.
In addition,
dApps have been proposed as micro-services for real-time inference ($\le\!10$\:ms) in the \glspl{cu}/\glspl{du}~\cite{doro2022dapps}.

%
The advantages of this cloud-based approach are: (i) it enables dynamic reconfiguration of the \ran by instantiating disaggregated
functionalities, xApps and rApps on-the-fly to meet current demand and requirements~\cite{schmidt2021ran,niknam2022intelligent,bonati2021intelligence}; and (ii) it
reduces the total cost of ownership via cloud infrastructure sharing (i.e., sharing of data centers, servers and network equipment)~\cite{ojiaghi2023benefits}. 


However, \ran cloudification comes with possible downsides. First, it expands the compute surface, thus potentially increasing the power consumption of the \ran. Second, implementing intelligent control via micro-services in a cloud environment (called O-Cloud in \oran) may not provide tight performance guarantees required to close the control loops in the real, near-real, or non-real time scales. While timing constraints of virtualized \glspl{ran} have been studied extensively in the literature with respect to the user plane~\cite{garciaaviles2021nuberu,garcia2018fluidran,parvez2018survey,giannone2019impact}, \textit{how to achieve the same guarantees in the control plane is still an open challenge, especially regarding control loops and decisions made by the \rics.} Guaranteeing such constraints in the control plane is necessary to ensure that such decisions are timely and do not become obsolete by the time they are enforced.

\begin{figure}[t]
\setlength\abovecaptionskip{1pt}
\ifexttikz
    \tikzsetnextfilename{mix}
\fi
\begin{subfigure}[t]{0.48\columnwidth}
    \setlength\abovecaptionskip{0pt}
    \centering
    \setlength\fwidth{0.7\columnwidth}
    \setlength\fheight{0.4\columnwidth}
    \includegraphics[width=\columnwidth]{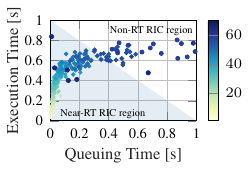}
    \caption{Execution time}
    \label{fig:mix-intro}
\end{subfigure}\hfill
\ifexttikz
    \tikzsetnextfilename{latency-stddev}
\fi
\begin{subfigure}[t]{0.48\columnwidth}
    \setlength\abovecaptionskip{0pt}
    \centering
    \setlength\fwidth{0.7\columnwidth}
    \setlength\fheight{0.4\columnwidth}
    \input{figures/latency-stddev.tex}
    \caption{Inference time}
    \label{fig:latency-stddev}
\end{subfigure}
\setlength\belowcaptionskip{-.6cm}
\caption{Left: execution vs.\ queuing times for different xApp number. The number of xApps is indicated by the color map. Right: inference time vs.\ number of xApps. Shaded areas represent \nearrt \ric $1$\:s threshold.}
\label{fig:intro}
\end{figure}

Indeed, poorly-managed O-Cloud environments for rApps, xApps, and dApps can easily lead to control deadline violations, as shown in Fig.~\ref{fig:intro}.
%
Fig.~\ref{fig:mix-intro} reports
(i) the queuing time, i.e., the time needed by the \nearrt \gls{ric} to de-queue input data from the \gls{ran}
and feed it to an xApp (x-axis); (ii) the execution time, i.e., the time needed by the xApp to process the input and generate an output (y-axis). Fig.~\ref{fig:latency-stddev} reports the inference time, i.e., the sum of queuing and execution time, with an increasing number of xApp (indicated by the colorbar in Fig.~\ref{fig:mix-intro}) executed on the \ric. This example is based on measurements taken on an \oran-compliant \nearrt \ric deployed on a Red Hat OpenShift cluster, where we instantiate xApps with diverse \ai workloads through the \gls{ci}/\gls{cd} pipelines provided by NeutRAN~\cite{bonati2023neutran}.
The goal is to close the loop within the $1$\:s \nearrt \ric region (shaded areas in Fig.~\ref{fig:intro}). In this case, the OpenShift fails to satisfy the control latency guarantees when the number of xApps exceeds 50, which is a conservative estimate
for most real-world applications with hundreds of base stations~\cite{polese2021machine}.
%
%
%

In the cloud industry, compute resource scaling is an established strategy to cope with the need for extra processing power, and is
a well-investigated topic in the literature~\cite{vaquero2011dynamically,bauer2019chamulteon,gulati2011cloud}, with a variety of approaches ranging from heuristic schedulers to predictive and ML-based models~\cite{singhvi2021atoll,hu2021k8s,rossi2019horizontal,casalicchio2017auto,ABENI2020101709,distefano2020ananke,prachitmutita,han2012lightweigth}. However, these solutions focus on ensuring that generic micro-services properly execute on the available compute resources, but \emph{do not provide performance guarantees on latency-critical applications}. 
As an example, in a widely used framework like Kubernetes scaling is obtained either by regulating the amount of resources allocated to each service, or by increasing the number of active worker nodes in the compute cluster. However, this approach
is based on resource utilization (e.g., CPU, RAM) and not on latency constraints (we will show that CPU/RAM-based scaling alone is unsuitable to ensure timely \ran control)~\cite{k8sscaling}. Previous work has also addressed scaling with deadline constraints, but it considers long-term or stochastic latency metrics and leverage heuristic solutions rather than optimization~\cite{singhvi2021atoll,mao2010cloud,mao2011auto,anagnostou2019towards,das2016automated}.
Moreover, uncontrolled and sub-optimal scaling might unnecessarily utilize
excessive resources, thus increasing the energy consumption and costs (capital and operational), making the \oran proposition less attractive for network operators~\cite{dryjanskiran}. Therefore, \emph{it is crucial to explore and understand this complex trade-off between latency and energy consumption.}


Our objective is to explore this trade-off and provide an optimization framework for scaling compute resources in the O-Cloud that is (i) aware of specific O-RAN application requirements; and (ii) satisfies inference constraints while minimizing energy consumption.
We make the following contributions: 

    
    
    
%
\noindent
\begin{itemize}[wide = 0pt]
    \item We propose \name, a tunable auto-scaling framework for \oran systems, capable of managing \ai-based xApps, rApps, and dApps on shared computing clusters with latency guarantees while considering important aspects such as profit and energy consumption (Secs.~\ref{section:framework} and~\ref{section:system_model}).
    
    \item We perform an extensive data collection campaign on the \gls{osc} \nearrt RIC deployed on an OpenShift cluster to evaluate how resource sharing and scaling affect inference times of \ai-based \oran applications (Sec.~\ref{sec:latency}). We leverage these measurements to derive a data-driven latency model that is used by \name to efficiently instantiate xApps, rApps, and dApps to satisfy application-specific latency requirements.
    
    \item We formulate the latency-constrained instantiation and scaling problem as a \gls{qcqp} that we prove NP-hard (Sec.~\ref{sec:optimization_model}). We solve the problem via branch-and-bound and evaluate \name's effectiveness via simulations (Sec.~\ref{section:numerical-evaluation}). Results show that scaling \oran applications is a time-constrained problem where congestion is measured on how fast \ai can produce outputs to guarantee continuous decision-making at diverse time scales, and not simply on how many resources are used. 
    
    \item We prototype \name as an rApp and perform an extensive experimental campaign on an \oran-compliant testbed. 
    Results show that \name can effectively perform instantiation and scaling tasks while guaranteeing desired application-specific latency requirements  (Sec.~\ref{sec:results}). 
\end{itemize}

\section{\textsc{Related Work}}
\label{section:related-work}%
Dynamic scaling of virtual 
machines or micro-services 
has been widely studied in the last decade~\cite{vaquero2011dynamically,bauer2019chamulteon}.
When considering scaling with latency guarantees, Singhvi et al. manage application latency with a deadline-aware scheduler in a server-less environment~\cite{singhvi2021atoll}. Mao et al. model virtual workload deadlines and costs, but for long-running applications rather than real, near-real, or non-real time control~\cite{mao2010cloud,mao2011auto}. Anagnostou et al. consider auto-scaling to meet deadlines for simulation workloads~\cite{anagnostou2019towards}. Das et al.~\cite{das2016automated} scale resources to meet query deadlines for a relational database, using a token bucket approach. Compared to prior work, in this paper we focus on tight control timelines, combine energy minimization or profit maximization, and scale resources solving a \gls{qcqp} based on a detailed model of \ran control workloads.

Open \ran is extending the cloud domain to cellular network functions. Several \gls{vnf} scaling solutions have been proposed, but without considering latency guarantees for closed-loop control~\cite{fei2018adaptive,8761272,9463941}. In the O-RAN context, Ali et al. analyze how to proactively scale resources for \glspl{vnf} with  workloads prediction~\cite{ali2023proactive}. D'Oro et al.\ orchestrate applications deployment, without however considering scaling or energy efficiency~\cite{doro2022orchestran}.
In the user plane, Garcia-Aviles et al.\ design a framework to preserve synchronization among base stations and users, maximize network throughput, and save resources in the presence of computing capacity shortages~\cite{garciaaviles2021nuberu}. Thaliath et al.~\cite{thaliath2022predictive} proactively scale resources to support network slices. However, these works are more concerned with optimally placing or executing services across the Open \gls{ran} infrastructure, rather than on guaranteeing control latency and minimizing energy consumption, as we do in this work.

Finally, energy efficiency is a priority for virtualized Open \ran. Prior literature work investigated energy consumption for the \gls{ran}---which consumes most of the energy in a cellular system~\cite{7864818}---as well as for \glspl{vnf} (e.g., core network,
\glspl{ric}). Ayala-Romero et al.\ optimize virtualized \gls{ran} power consumption, evaluating waveform trade-offs in different signal-to-noise ratio regimes~\cite{experimental-pw-vbs}. Pamuklu et al.\ propose a mixed linear programming problem for energy optimization, mindful of maximum tolerable delays for the data plane of the \gls{ran}~\cite{joint-opt-energy-latency}. Bonati et al.\ minimize \ran power consumption with dynamic power control orchestrated by a centralized controller~\cite{bonati2020cellos}.
Compared to these works, and to the best of our knowledge, \name is the first framework that optimally combines compute scaling, energy minimization, and timing constraints for \ran control in \oran, including an experimental inference characterization for different control workloads, and an experimental prototype.
%
%
%
%


%
%
%
%
%
%



\section{\name Architecture and Prototype}
\label{section:framework}

\begin{figure}[t!]
    \setlength\belowcaptionskip{-0.4cm}
    \centering
    \includegraphics[width=0.95\columnwidth]{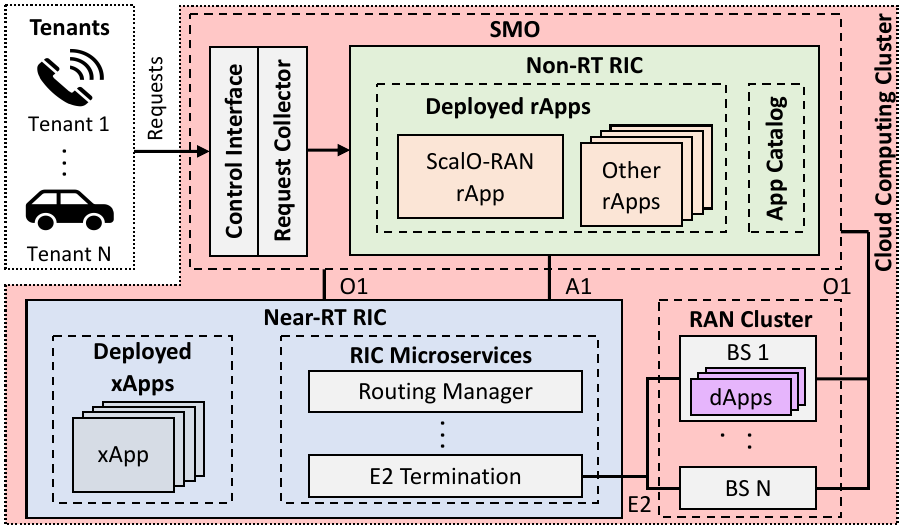}
    \caption{\name within the \oran architecture.}
    \label{fig:architecture}
\end{figure}

Fig.~\ref{fig:architecture} illustrates \name and its integration with the \oran architecture. 
A set of tenants (e.g., network operators) interface with a \textit{control interface} in the \gls{smo} to submit their request to deploy \ai-based \oran applications.
Requests are collected by a \textit{request collector} hosted in the \gls{smo}, and forwarded to \name rApp every $T$ seconds.
$T$ is a tunable parameter large enough to account for the time needed by \name \textit{optimization engine} to compute a solution, and for the time needed to instantiate the requested rApps, xApps and dApps.
%
%
After receiving these requests, \name computes an optimal instantiation and scaling policy to accommodate them, while making sure that demand and temporal constraints are satisfied (see Sec.~\ref{sec:optimization_model} for details on the optimization process).
%
Then, rApps, xApps and dApps are instantiated from the \textit{app catalog} to the selected servers---i.e., on the servers running the \nonrt \gls{ric} for rApps, \nearrt \gls{ric} for xApp, and \gls{cu}/\gls{du} for dApps
---according to the optimal solution found in the previous step.
%

\begin{figure}[t!]
    \setlength\belowcaptionskip{-0.4cm}
    \centering
    \includegraphics[width=.9\columnwidth]{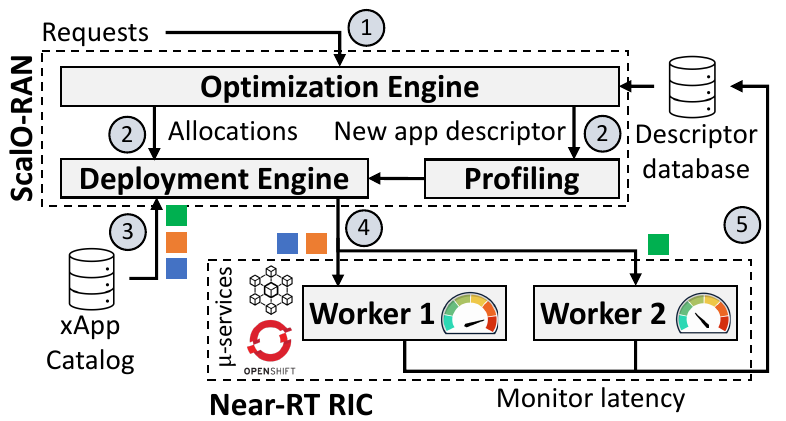}
    \caption{\name OpenShift-based prototype.}
    \label{figure:prototype}
\end{figure}

\textbf{\name Prototype.} We prototype \name on a Red Hat OpenShift cluster with 8~Dell PowerEdge servers, i.e., 3~control nodes and 5~worker nodes---two of which are reserved for \name workloads---running various Open \gls{ran} components, e.g., \gls{osc} \glspl{ric}, Open5GS core network, and cellular base stations based on srsRAN and OpenAirInterface.
Fig.~\ref{figure:prototype} depicts the main building blocks of our prototype, which implements \name procedures in steps 1-5 building on the \gls{ci}/\gls{cd} pipelines provided by NeutRAN~\cite{bonati2023neutran}. The prototype enables
latency profiling for xApps and embeds \name as an rApp to optimize workloads deployment.
Although \name is general, our prototype focuses on xApp instantiation.

First, requests from the tenants to deploy xApps are received by the \gls{smo}, and forwarded to the \name optimization engine (Step~1).
%
Each xApp is stored in a \textit{App Catalog} and is assigned an \textit{app descriptor} (Sec.~\ref{section:system_model}) that specifies type, objective, input/output format of the embedded \ai, among others. 
%
Upon receiving a request, \name controls whether the requested xApps are present in the catalog. 
%
New xApps (which lack of an app descriptor) are first profiled to benchmark their performance requirements (Step~2a),
by deploying the xApp on an idle worker through \name deployment engine. xApp deployment on the \nearrt \ric is automated using the \texttt{dms\_cli} tool and Helm charts~\cite{near-rt-ric-installation}. 
In case of xApps with an app descriptor, the optimization engine computes the optimal xApp allocation policy (using MATLAB and Gurobi) to satisfy the received requests, and forwards the result to the deployment engine (Step~2b).
The latter retrieves the xApps to instantiate from the xApp catalog (Step~3), and allocates them on the available worker nodes (e.g., worker~1 and worker~2 in the figure), based on the xApp latency constraints and on the expected run-time profile of the node (Step~4).
%
%
%
Finally, the nodes of the cluster periodically report their run-time latency to \name control interface (Step~5).

\section{{System Model}}
\label{section:system_model}%


In this section, we introduce our system model and notation.

\textbf{Infrastructure.}
We consider the Open \gls{ran} architecture proposed by the \oran Alliance~\cite{polese2022understanding}.
%
%
The cloud infrastructure is represented by the O-Cloud, which hosts \ran functions (i.e., \glspl{cu}, \glspl{du}, and \glspl{ru}), the \nonrt and \nearrt \glspl{ric}, xApps, rApps, and the \gls{smo} framework.
%
%
%
%
The O-Cloud computing infrastructure has access to a set $\servers$ of $S=|\servers|$ servers.
Although in principle servers in $\servers$ could also host \ran functions and \rics (see Fig.~\ref{fig:architecture}), data-driven \oran applications consume significant resources (e.g., CPU, RAM), and they might congest the server where they execute (see Fig.~\ref{fig:intro}).
%
To ensure reliability and availability of networking functionalities, we assume that rApps, xApps, and dApps execute on dedicated servers and let $\servers$ denote their set only. We also identify with $\servers^{\mathrm{CU/DU}}\subseteq\servers$ the servers co-located with a \gls{cu}/\gls{du} that can host dApps.
We design \name to be in charge of instantiating applications and scaling computing resources for a single cluster. In the case of $C$ clusters, $C$ instances of \name can be instantiated to serve each individual cluster. 

\textbf{\oran applications.} The rApps, xApps, and dApps available to the tenants are stored in a catalog $\apps$ on the \nonrt \gls{ric}, with $A=|\mathcal{A}|$ \ai-based applications. Without loss of generality, $\apps = \apps^{\mathrm{rApp}} \cup \apps^{\mathrm{xApp}} \cup \apps^{\mathrm{dApp}}$. Each application $a\in\apps$ is described via an \textit{app descriptor} that specifies the delivered functionality
(e.g., \ran slicing, traffic steering),
the type of \ai used (e.g., \gls{drl}, \gls{lstm}, \gls{cnn}), the type of application (e.g., xApp, rApp, dApp), the format and shape of input and output data (e.g., list of input \kpis and their shape, as well as type of action performed and its format), and its latency profile as detailed in Sec.~\ref{sec:latency}.

\textbf{Requests.} Tenants
sharing the \oran infrastructure might have conflicting interest, different business goals, and serve users
with different \glspl{sla}. To satisfy these requirements and meet their goals, tenants submit requests to deploy a selection of rApps, xApps, and dApps from the catalog $\apps$. 
Let $\reqs$ be the set of requests submitted by all tenants. A request is modeled as a tuple $r=(\mathbf{n}_r,\mathbf{L}_r,\boldsymbol{\delta}_r)$ where $\mathbf{n}_r=(n_{r,a})_{(r,a)\in\reqs \times \apps}$, $\mathbf{L}_r=(L_{r,a})_{(r,a)\in\reqs \times \apps}$, $\boldsymbol{\delta}_r=(\delta_{r,a,s})_{(r,a,s)\in\reqs \times \apps \times \servers}$, and $\times$ indicates the Cartesian product. $n_{r,a}$ represents the number of applications of type $a\in\apps$ that need to be instantiated to satisfy request $r$. Similarly, $L_{r,a}$ represents the maximum inference time that the tenant tolerates executing applications of type $a$ on any server. 
For example, a tenant could request $n_{r,a'}=4$ xApps to control \ran slicing policies of 4 \glspl{du} at a maximum tolerable inference time of $L_{r,a'}=100$\:ms, as well as $n_{r,a''}=1$ rApp to control handover management with a desired inference time of $L_{r,a''}=10$\:s. 
Note that controlling several \ran components with a single xApp/rApp is generally to be avoided, as it might result in congestion and large inference times.
%
Hence, we assume $n_{r,a}\geq1$. 

Tenants might submit requests that do not require any maximum inference time guarantee (e.g., $L_{r,a}=+\infty$). However, by design the \nearrt \ric should take decisions within $1$~s, while dApps should take decisions within $10$\:ms. For this reason, we also introduce a requirement $L^{\mathrm{APP}}_a$ that ensures that any application of type $a$ produces an
output within $L^{\mathrm{APP}}_a$. For example, if application $a\in\apps^{\mathrm{xApp}}$, $L^{\mathrm{APP}}_a=1$~s, while $L^{\mathrm{APP}}_a=10$\:ms if $a\in\apps^{\mathrm{dApp}}$. Since \oran specifications do not provide any maximum inference time requirements for rApps, we set $L^{\mathrm{APP}}_a=+\infty$ for $a\in\apps^{\mathrm{rApp}}$.

We also introduce the parameter $\delta_{r,a,s}\in\{0,1\}$ to identify the execution location of dApps. Specifically, $\delta_{r,a,s}=1$ indicates that a dApp $a\in\apps^{\mathrm{dApp}}$ needs to be executed at server $s$ co-located with a \gls{cu}/\gls{du}. Since $s$ unequivocally identifies each server, we can use $s$ to identify the target \gls{cu}/\gls{du} required by the tenant. We set $\delta_{r,a,s}=0$ for all $a\in\apps\setminus\apps^{\mathrm{dApp}}$ and servers. 

\subsection{Notation and variables} 
\label{sec:notation}


We introduce the server activation profile $\mathbf{x}=(x_s)_{s\in\servers}$, where $x_s\in\{0,1\}$ indicates whether server $s$ is actively hosting at least one \ai-based \oran application ($x_s=1$) or not ($x_s=0$). 
To capture the allocation and instantiation of applications across the different servers, 
we introduce an allocation variable $\mathbf{y}=(y_{r,a,s})_{(r,a,s) \in \reqs \times \apps \times \servers}$ that indicates how many instances of app $a$ for request $r$ have been instantiated on server $s$. For each request $r$ and application $a$, the variables $y_{r,a,s}$ are defined over the $(A\cdot S-1)$-simplex $\Delta_{r,a}=\{(y_{r,a,s})_{s \in \servers}, y_{r,a,s} \in \mathbb{Z}^+_0 | \sum_{s\in\servers} y_{r,a,s} = n_{r,a}\}$, with $\mathbb{Z}^+_0$ being the set of positive integer numbers including $0$. It follows that $x_s = 1$ if and only if $\sum_{a\in\apps} \sum_{r\in\reqs} y_{r,a,s} > 1$. We introduce an auxiliary indicator variable $w_{r,s}\in\{0,1\}$ for all $r\in\reqs$ and $s\in\servers$ such that $w_{r,s}=1$ if and only if $\sum_{a\in\apps} y_{r,a,s} > 0$, i.e., server $s$ is hosting at least one instance of any application required by request $r$.

We also introduce an indicator variable $z_r\in\{0,1\}$ that, for each $r\in\reqs$, represents whether the allocation variable $\mathbf{y}$ satisfies the requirements of request $r$, both in terms of
instances to be deployed, as well as latency
($z_r=1$), or not ($z_r=0$).
We also define the indicator variable $\pi_{a,s}\in\{0,1\}$ to determine the number $A_s$ of different applications that have at least one instance running on server $s$. For all $a\in\apps$ and $s\in\servers$, $\pi_{a,s}=1$ if server $s$ has at least one instance of application $a$, i.e., $\sum_{r\in\reqs} y_{r,a,s} > 0$, and $\pi_{a,s}=0$ otherwise. $A_s$ is defined~as:
\begin{equation} \label{eq:w_as}
    A_s = \sum_{a} {\pi_{a,s}}. 
    \vspace{-0.2cm}
\end{equation}

%
Finally, we define the following variables, $\mathbf{z}=(z_r)_{r\in\reqs}$, $\mathbf{w}=(w_{r,s})_{(r,s)\in\reqs \times \servers}$, and $\boldsymbol{\pi}=(\pi_{a,s})_{(a,s)\in\apps \times \servers}$.



\section{Inference time of \oran Applications}
\label{sec:latency}%

To properly satisfy inference constraints, we first derive a latency model to regulate scaling and instantiation procedures and ensure that all applications can close the control loop within the desired temporal window. 
This section reports the results of a data collection campaign, where we leverage the OpenShift \name prototype (Sec.~\ref{section:framework}) to gather data on how congestion and resource sharing affect the inference time of different \ai architectures and algorithms. 

\subsection{Profiling inference time}


The inference time of \ai-based \oran applications heavily depends on the complexity of the \ai algorithms and architectures embedded in dApps, xApps, and rApps (e.g., width, depth, number of parameters and layers, need for convolutions). 
Moreover, as shown in Fig.~\ref{fig:intro}, the more applications coexist on the same hardware and share its resources, the more the inference time increases due to constrained computational resources.
Thus, to properly quantify how resource sharing of coexisting applications affects their inference time, it is imperative to derive a model capable of capturing such dynamics.

\ai for \oran systems can perform classification (e.g., anomaly detection), forecasting (e.g., \kpi prediction), and control (e.g., resource allocation)~\cite{brik2022deep}. Even if these tasks can be performed with multiple \ai architectures (e.g., classification can use \glspl{cnn}, Decision Trees), in our analysis we consider three well-established and diverse \ai models for each of the above tasks. Specifically, for \textit{classification}, we use a \gls{cnn} with 231,875 parameters and a fully connected output layer;
for \textit{forecasting} we use a \gls{lstm} with 49,987 parameters, bidirectional memory cells, and a fully connected output layer; and for control we use a \gls{drl} agent with more than 50,000 parameters.

Our goal is to derive an inference time model to scale intelligent \oran applications. Thus, we only focus on evaluating their inference time, which is the same whether the \ai has been trained or not, as the number of operations (e.g., multiplications, convolutions, additions) to perform is the same.

We consider a single worker node of the OpenShift cluster and 
%
we deploy one xApp instance at a time.
To collect the data at scale, we developed an E2 traffic generator
using the open-source \oran dataset from~\cite{polese2022coloran}. The generator emulates E2 traffic by constantly extracting at random \kpis from the dataset, with a format that matches the input expected by the xApp \ai models (e.g., which \kpi to extract and the shape of the input), as specified in the \textit{app descriptor} of each xApp. 

Whenever we add a new instance of xApp $a$ on server $s$, we use the traffic generator to produce input data for the new instance and
measure three types of latency: (i)~\textit{queuing time} $t^{\mathrm{queue}}_{a,s}$, which measures how long it takes for the xApp to ingest the input once it has been received at the E2 termination of the \nearrt \ric; (ii)~\textit{execution time} $t^{\mathrm{exec}}_{a,s}$, measuring the time to produce an output once an xApp  receives an input; and (iii)~\textit{inference time} $t^{\mathrm{inf}}_{a,s} = t^{\mathrm{queue}}_{a,s} + t^{\mathrm{exec}}_{a,s}$. We also keep track of CPU and RAM utilization of the server.

\begin{figure}[t!]
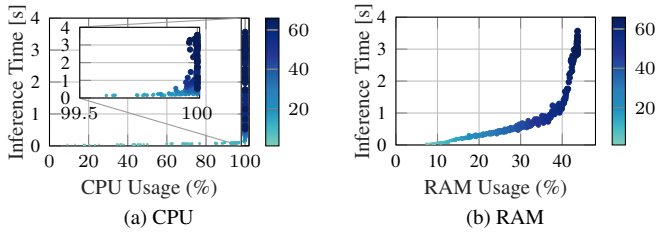

\setlength\abovecaptionskip{1pt}
\ifexttikz
    \tikzsetnextfilename{cpu-example}
\fi
\begin{subfigure}[t]{0.48\columnwidth}
    \setlength\abovecaptionskip{0pt}
    \centering
    \setlength\fwidth{0.75\columnwidth}
    \setlength\fheight{0.4\columnwidth}
    \input{figures/cpu-example.tex}
    \caption{CPU}
    \label{fig:cpu}
\end{subfigure}\hfill
\ifexttikz
    \tikzsetnextfilename{ram-example}
\fi
\begin{subfigure}[t]{0.48\columnwidth}
    \setlength\abovecaptionskip{0pt}
    \centering
    \setlength\fwidth{0.75\columnwidth}
    \setlength\fheight{0.4\columnwidth}
    \input{figures/ram-example.tex}
    \caption{RAM}
    \label{fig:ram}
\end{subfigure}
\caption{Inference time vs.\ CPU and RAM usage for different number of xApps. The color map indicates the number of xApps.}
\label{fig:cpu_ram_example}
\vspace{-0.3cm}
\end{figure}

One could also consider both the time needed to forward the \kpis and the control action between the \ran and the \rics. However, since all servers are co-located in the same cluster, these parameters are constant. Moreover, data over high-speed optical fiber links has low and predictable latency (few hundreds of milliseconds, including switching), which is negligible if compared to the times cale of the \nearrt \ric (i.e., below 1~s) and \nonrt \ric (i.e., above 1~s). For these reasons, we do not include these terms in our model.
Under these assumptions, we define the inference time when $y$ instances of application $a\in\apps$ are executing on server $s$ as follows:
\begin{equation} \label{eq:interence_latency}
    t^{\mathrm{inf}}_{a,s}(y) =  t^{\mathrm{exec}}_{a,s}(y) + t^{\mathrm{queue}}_{a,s}(y).
\end{equation}

\subsection{Deriving a latency model}



\begin{figure}[tb]
\setlength\abovecaptionskip{5pt}
\ifexttikz
    \tikzsetnextfilename{legend-inf-latency}
\fi
\begin{subfigure}[t]{\columnwidth}
    \centering
    \setlength\fwidth{0.7\columnwidth}
    \setlength\fheight{0\columnwidth}
%
%
\begin{tikzpicture}
\pgfplotsset{
every tick label/.append style={font=\scriptsize},
colormap/YlGnBu
}

\begin{axis}[%
width=0.951\fwidth,
height=\fheight,
at={(0\fwidth,0\fheight)},
scale only axis,
xmin=0,
xmax=120,
xlabel={Number of xApp},
ymin=0,
ymax=1400,
yticklabels={,0,0.5,1},
ylabel={Inference Time [s]},
axis background/.style={fill=white},
xmajorgrids,
ymajorgrids,
legend style={legend cell align=left, align=left, draw=white!15!black,
anchor=north west, font=\scriptsize, align=left, at={(0.02,0.98)}},
%
legend columns=3,
xlabel style={font=\footnotesize\color{white!15!black}},
ylabel style={font=\footnotesize\color{white!15!black}},
yticklabel shift=-2pt,
ylabel shift=-5pt,
xlabel shift=-2pt,
clip mode=individual,  
hide axis
]
\addlegendimage{color=RdBu-J, only marks, mark=*, mark size=1pt, mark repeat=5, mark options={solid, RdBu-J}}
\addlegendentry{Measured};

\addlegendimage{color=RdBu-C, thick}
\addlegendentry{Average fit};

\addlegendimage{color=RdBu-O, very thick, dashed}
\addlegendentry{Conservative fit};

\end{axis}

\end{tikzpicture}%
\end{subfigure}
\vspace{-.15cm}
\ifexttikz
    \tikzsetnextfilename{cnn}
\fi
\begin{subfigure}[t]{0.48\columnwidth}
    \setlength\abovecaptionskip{0pt}
    \centering
    \setlength\fwidth{0.7\columnwidth}
    \setlength\fheight{0.4\columnwidth}
    \includegraphics[width=\textwidth]{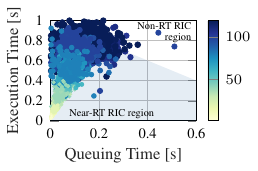}
    \caption{CNN execution time}
    \label{fig:cnn}
\end{subfigure}\hfill
\ifexttikz
    \tikzsetnextfilename{cnn-inf-latency}
\fi
\begin{subfigure}[t]{0.48\columnwidth}
    \setlength\abovecaptionskip{0pt}
    \centering
    \setlength\fwidth{0.7\columnwidth}
    \setlength\fheight{0.4\columnwidth}
    \input{figures/cnn-inf-latency.tex}
    \caption{CNN inference time}
    \label{fig:latency-cnn}
\end{subfigure}
\vspace{-.15cm}
\ifexttikz
    \tikzsetnextfilename{lstm}
\fi
\begin{subfigure}[t]{0.48\columnwidth}
    \setlength\abovecaptionskip{0pt}
    \centering
    \setlength\fwidth{0.7\columnwidth}
    \setlength\fheight{0.4\columnwidth}
    \includegraphics[width=\textwidth]{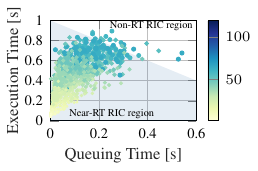}
    \caption{LSTM execution time}
    \label{fig:lstm}
\end{subfigure}\hfill
\ifexttikz
    \tikzsetnextfilename{lstm-inf-latency}
\fi
\begin{subfigure}[t]{0.48\columnwidth}
    \setlength\abovecaptionskip{0pt}
    \centering
    \setlength\fwidth{0.7\columnwidth}
    \setlength\fheight{0.4\columnwidth}
    \input{figures/lstm-inf-latency.tex}
    \caption{LSTM inference time}
    \label{fig:latency-lstm}
\end{subfigure}
\vspace{-.15cm}
\ifexttikz
    \tikzsetnextfilename{drl}
\fi
\begin{subfigure}[t]{0.48\columnwidth}
    \setlength\abovecaptionskip{0pt}
    \centering
    \setlength\fwidth{0.7\columnwidth}
    \setlength\fheight{0.4\columnwidth}
    \includegraphics[width=\textwidth]{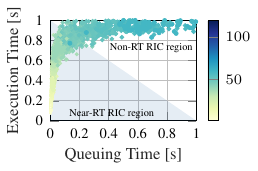}
    \caption{DRL execution time}
    \label{fig:drl}
\end{subfigure}\hfill
\ifexttikz
    \tikzsetnextfilename{drl-inf-latency}
\fi
\begin{subfigure}[t]{0.48\columnwidth}
    \setlength\abovecaptionskip{0pt}
    \centering
    \setlength\fwidth{0.7\columnwidth}
    \setlength\fheight{0.4\columnwidth}
    \hspace{6.5pt}\input{figures/drl-inf-latency.tex}
    \caption{DRL inference time}
    \label{fig:latency-drl}
\end{subfigure}
\caption{Left: execution time vs.\ queuing time for different number of xApps. The color map indicates the number of xApps. Right: inference time vs.\ number of xApps. $\otimes$ represents the break point of the piecewise linearization functions.}
\label{fig:scatter-plot-single-xapps}
\vspace{-0.2cm}
\end{figure}

In Fig.~\ref{fig:cpu_ram_example}, we show how the inference time varies as a function of the CPU utilization and number of xApps (uniformly distributed between \gls{cnn}, \gls{lstm} and \gls{drl}). We notice that 20 xApps already consume 100\% of the CPU: this saturation prevents us from accurately modeling the execution time from the CPU utilization alone. RAM usage brings more insights, but predicting inference time from RAM occupation is hard as two models might use the same RAM but execute at different speeds.
To overcome the above limitation, we instead focus on measuring both $t^{\mathrm{exec}}_{a,s}$ and $t^{\mathrm{queue}}_{a,s}$ and derive an inference time model from these parameters. 
%

%
To better understand how execution and queuing time affect $t^{\mathrm{inf}}_{a,s}(y)$, in Figs.~\ref{fig:cnn}, \ref{fig:lstm}, and~\ref{fig:drl} we show $t^{\mathrm{exec}}(y)$ and $t^{\mathrm{queue}}(y)$ for the different xApp types when $y$ instances of the same xApp execute on the server, while we show the respective $t^{\mathrm{inf}}_{a,s}(y)$ in Figs.~\ref{fig:latency-cnn}, \ref{fig:latency-lstm}, and~\ref{fig:latency-drl}. 
Instead, in Fig.~\ref{fig:intro} we present results obtained by instantiating $y$ instances of the three xApps at the same time. 
The figures also show the regions identifying the \nearrt and \nonrt \rics operational domains. In general, we notice that the execution time $t^{\mathrm{exec}}_{a,s}(y)$ strongly affect inference time when $y$ is small, while $t^{\mathrm{queue}}_{a,s}(y)$ becomes relevant when $y$ grows due to congestion.
These results suggest that inference time can be modeled using an increasing function with two distinct regions: a region where inference time grows at a moderate rate with the number of applications running on the server, and a congestion region with a steep increasing trend.

Although one can compute such functions in several ways (e.g., linear regression, neural networks), we aim at estimating latency with a model that is accurate, simple to integrate into an optimization problem, and reduces the underestimation risks to avoid deploying \ai that would violate maximum latency requirements. For this reason, we model inference time via piecewise linear regression. This has several advantages: it is general, it can be used to accurately approximate non-linear functions, and can be used to remove non-linearities in optimization problems, thus resulting in lower complexity~\cite{dunham1986optimum}.
In general, one could compute the minimum amount of segments necessary for the approximation by using the piecewise linearization methods in \cite{dunham1986optimum}. 
However, our data analysis suggests that inference time behaves as an ``elbow'' function. We thus use 2-segment piecewise linear regression~\cite{vieth1989fitting}, which describes a function $f(y)$ as $f(y)=\lambda_1 \cdot y + b_1$ if $y\leq y_0$, and $f(y)=\lambda_2 \cdot y + b_2$ if $y>y_0$, where $y_0$ is the break point, $\lambda_i$ is the slope and $b_i$ is the intercept of the $i$-th segment.

Figs.~\ref{fig:latency-cnn}, \ref{fig:latency-lstm}, and~\ref{fig:latency-drl} also show that $t^{\mathrm{inf}}_{a,s}(y)$ can be approximated using the following 2-segment piecewise linear function
\begin{align} \label{eq:piecewise}
    \tilde{t}^{\mathrm{inf}}_{a,s}(y) = & \begin{cases} 
                        \lambda^I_{a,s}\cdot y + b^I_{a,s} \hspace{2cm} \mathrm{if}~y \leq \tilde{y}_{a,s}  \\
                        \lambda^{II}_{a,s} \cdot y + b^{II}_{a,s} \hspace{2cm} \mathrm{otherwise},
                       \end{cases}
\end{align}
\noindent
where $\tilde{y}_{a,s}$ is the break point, and $\lambda^i_{a,s}$ and $b^i_{a,s}$ are the slope and intercept of the $i$-th segment, with $i\in\{I,II\}$. 
%
The values of $\tilde{y}_{a,s}$, $\lambda^I_{a,s}$m and $\lambda^{II}_{a,s}$ for the applications considered in our campaign are extracted via piecewise linear regression from the data collected on our prototype and are reported in Table~\ref{table:piecewise}.

\begin{table}[t]
\setlength\abovecaptionskip{1pt}
\caption{Piecewise Regression Parameters.}
\label{table:piecewise}
\renewcommand{\arraystretch}{1.3}
\addtolength{\tabcolsep}{-0.45em}
\centering
\begin{tabular}{r|ccccc|ccccc|}
\cline{2-11}
\multicolumn{1}{l|}{} & \multicolumn{5}{c|}{Conservative fit} & \multicolumn{5}{c|}{Average fit} \\ \cline{2-11} 
\multicolumn{1}{l|}{} & \multicolumn{1}{c|}{$\lambda^I$} & \multicolumn{1}{c|}{$b^I$} & \multicolumn{1}{c|}{$\lambda^{II}$} & \multicolumn{1}{c|}{$b^{II}$} & $\tilde{y}$ & \multicolumn{1}{c|}{$\lambda^I$} & \multicolumn{1}{c|}{$b^I$} & \multicolumn{1}{c|}{$\lambda^{II}$} & \multicolumn{1}{c|}{$b^{II}$} & $\tilde{y}$ \\ \hline
\multicolumn{1}{|r|}{CNN} & \multicolumn{1}{c|}{9.057} & \multicolumn{1}{c|}{18.94} & \multicolumn{1}{c|}{11.73} & \multicolumn{1}{c|}{-218.9} & 92 & \multicolumn{1}{c|}{1.535} & \multicolumn{1}{c|}{20.97} & \multicolumn{1}{c|}{8.237} & \multicolumn{1}{c|}{-22.3} & 9 \\ \hline
\multicolumn{1}{|r|}{LSTM} & \multicolumn{1}{c|}{17.27} & \multicolumn{1}{c|}{32.73} & \multicolumn{1}{c|}{18.21} & \multicolumn{1}{c|}{-10.68} & 49 & \multicolumn{1}{c|}{3.498} & \multicolumn{1}{c|}{38.99} & \multicolumn{1}{c|}{15.26} & \multicolumn{1}{c|}{-43.47} & 9 \\ \hline
\multicolumn{1}{|r|}{DRL} & \multicolumn{1}{c|}{24.88} & \multicolumn{1}{c|}{25.12} & \multicolumn{1}{c|}{130} & \multicolumn{1}{c|}{-5336} & 51 & \multicolumn{1}{c|}{20.56} & \multicolumn{1}{c|}{-10.54} & \multicolumn{1}{c|}{67.45} & \multicolumn{1}{c|}{-2250} & 48 \\ \hline
\end{tabular}
\vspace{-0.2cm}
\end{table}


The figures illustrate the outcome of piecewise linearization of the inference time function for the \gls{ml}-based control xApps for two cases: an \textit{average fit} where we approximate the average behavior; and a \textit{conservative fit} where we can account for upper bounds in the data via piecewise linear bounding~\cite{ngueveu2019piecewise}.
We note that both linearization offers a good approximation that captures the elbow-shaped behavior of the distribution.  
The average fit might result in underestimations and violation of latency requirements, as it only captures the expected behavior. To mitigate this phenomenon, we can use the conservative fit which also accounts for the variance of measurements, especially when the number of deployed applications is high. 

We now extend the application-specific inference model to a more general case where the same server hosts several instances of different applications. Note that while the measurement campaign profiled \ai models packaged as xApps, the same latency model would hold when they are packaged as dApps or rApps. Let $y_{r,a,s}$ be the number of instances of applications of type $a$ from any request $r$ executing on server $s$. The inference time of all instances executing on server $s$ can be expressed as
\begin{equation}\label{eq:latency_f_different}
\begin{aligned}
    l_{s}(\mathbf{y},\boldsymbol{\pi}) = \frac{1}{A_s} \sum_{a \in \apps}   \tilde{t}^{\mathrm{inf}}_{a,s}\left(Y_s\right) \pi_{a,s},
\end{aligned}
\end{equation}
\noindent
where $Y_s = \sum_{r\in\reqs} \sum_{a\in\apps} y_{r,a,s}$ is the total number of application instances hosted on $s\in\servers$, $\tilde{t}^{\mathrm{inf}}_{a,s}(\cdot)$ is defined in \eqref{eq:piecewise}, and $A_s$ from \eqref{eq:w_as} is a function of $\boldsymbol{\pi}$.
%
%
The expression in \eqref{eq:latency_f_different} models the expected value of the inference time when multiple instances of different applications are executed on the same server.



\section{\name Optimization Engine}
\label{sec:optimization_model}%

In this section, we first introduce the instantiation and scaling problem for intelligent \oran applications. Then, we discuss how to design an objective function that can capture diverse needs such as reducing energy and maximizing profit.

\subsection{Formulating the problem}

With the notation and variables defined in Sec.~\ref{sec:notation}, we are ready to formulate the Instantiation and Scaling Problem (ISP):

\vspace{-10pt}
\begin{align}
    \max_{\substack{\mathbf{x}, \mathbf{y}, \mathbf{z}, \\ \mathbf{w}, \boldsymbol{\pi}}} & \hspace{0.2cm}U(\mathbf{x}, \mathbf{y}, \mathbf{z}, \mathbf{w}, \boldsymbol{\pi}) \label{eq:problem} \tag{ISP} \\
    \mathrm{s.t.:} 
    &  \sum_{s\in\servers} y_{r,a,s} \geq n_{r,a} - M_1(1-z_r), \hspace{0.2cm} \forall (r,a)\in\reqs\times\apps \label{eq:con1a} \\
    &  \sum_{s\in\servers} y_{r,a,s} \leq n_{r,a} + M_1(1-z_r), \hspace{0.2cm} \forall (r,a)\in\reqs\times\apps \label{eq:con1b} \\
    &  y_{r,a,s} \leq n_{r,a} \cdot x_{s}, \hspace{1.37cm} \forall (r,a,s)\in\reqs\times\apps\times\servers \label{eq:con2} \\
    & l_{s}(\mathbf{y,\boldsymbol{\pi}}) w_{r,s} \leq L^\mathrm{MAX}_{r,a}, \hspace{0.6cm} \forall (r,a,s)\in\reqs\times\apps\times\servers \label{eq:con3} \\
    & \pi_{a,s} \leq x_s, \hspace{3.4cm} \forall (a,s)\in\apps\times\servers \label{eq:con4} \\
    & \sum_{r\in\reqs} {y_{r,a,s}} \geq 1 - M_2(1-\pi_{s,a}), \hspace{0.15cm} \forall (a,s)\in\apps\times\servers \label{eq:con5a} \\
    & \sum_{r\in\reqs} {y_{r,a,s}} \leq M_2\cdot\pi_{s,a}, \hspace{1.4cm} \forall (a,s)\in\apps\times\servers\label{eq:con5b}\\
    & \sum_{a\in\apps} y_{r,a,s} \geq 1 - M_3(1-w_{r,s}), \hspace{0.17cm} \forall (r,s)\in\reqs\times\servers \label{eq:con6a} \\
    & \sum_{a\in\apps} y_{r,a,s} \leq M_3\cdot w_{r,s}, \hspace{1.4cm} \forall (r,s)\in\reqs\times\servers \label{eq:con6b} \\
    & w_{r,s} \geq -M_4(1-x_s), \hspace{1.6cm} \forall (r,s)\in\reqs\times\servers \label{eq:con7a} \\
    & w_{r,s} \leq M_4\cdot x_s, \hspace{2.5cm} \forall (r,s)\in\reqs\times\servers \label{eq:con7b}\\
    & w_{r,s} \leq z_r, \hspace{3.3cm} \forall (r,s)\in\reqs\times\servers  \label{eq:con8} \\
    & y_{r,a,s} \leq \delta_{r,a,s}, \hspace{2.7cm} \forall (r,s)\in\reqs\times\servers  \label{eq:con9}
\end{align}
\noindent
where $U(\cdot)$ is the objective function (which we will discuss in Sec.~\ref{sec:objective}), $L^\mathrm{MAX}_{r,a}=\min\{L_{r,a},L^\mathrm{APP}_{a}\}$, $l_s(\mathbf{y},\boldsymbol{\pi})$ is defined in \eqref{eq:latency_f_different}, and $M_1 = M_2 = \sum_{a\in\apps} n_{r,a}+1$, $M_3 = \max_{(r,a)\in\reqs\times\apps}\{n_{r,a}\}+1$, $M_4=1$ are coefficients we use to formulate conditional constraints (e.g., applications can be instantiated on a server if and only if the server is active) using the big-M notation.
Specifically, \eqref{eq:con1a}-\eqref{eq:con1b} ensure that a sufficient condition for a request $r$ to be considered satisfied is that we must allocate all required applications and satisfy $\mathbf{n}_r$. \eqref{eq:con2} ensures that instances of application $a$ requested by $r$ can be instantiated only on active servers and the number of instances cannot exceed the demand. \eqref{eq:con3} ensures that all latency requirements (from tenants or from \oran specifications) are satisfied. \eqref{eq:con4} ensures that application instances can run on active servers only, while~\eqref{eq:con5a}-\eqref{eq:con5b} ensure that the indicator variable $\pi_{a,s}$ is activated if and only if there is at least one instance of application of type $a$ running on server $s$. Similarly, \eqref{eq:con6a}-\eqref{eq:con7b} ensure that $w_{r,s}=1$ if and only if there is at least one instance of any application requested by $r$ running on an active server $s$. Finally, \eqref{eq:con8} ensures that $w_{r,s}=1$ only if the request can be satisfied completely (i.e., if $z_r=1$), and~\eqref{eq:con9} guarantees that dApps are instantiated only at \glspl{cu} and \glspl{du} selected by the tenants. 
From~\eqref{eq:piecewise} and \eqref{eq:w_as}, \eqref{eq:con3} is non-linear but can be reformulated via the following big-M formulation
\begin{align} \label{eq:t_inf_con}
    \sum_{a\in\apps} \tilde{t}^{\mathrm{inf}}_{a,s}(Y_s) \pi_{a,s} \leq L^\mathrm{MAX}_{r,a} + M_5 (1-w_{r,s}),
\end{align}
\noindent
where $M_5$ is a large real-valued positive number, and $\tilde{t}^{\mathrm{inf}}_{a,s}(Y_s)$ is a piecewise function from \eqref{eq:piecewise} as follows
\begin{align} \label{eq:t_inf_piecewise}
    \tilde{t}^{\mathrm{inf}}_{a,s}(Y_s) & = \nu_{a,s} (\lambda^I_{a,s} \cdot  \sum_{r'\in\reqs}{y_{r',a,s}} + b^I_{a,s})  \nonumber \\ 
    & + (1-\nu_{a,s}) \left(\lambda^I_{a,s}\cdot \tilde{y}_{a,s} + \lambda^{II}_{a,s} \sum_{r'\in\reqs}{y_{r',a,s}}\right)
\end{align}
\noindent
where $\nu_{a,s}\in\{0,1\}$ is an auxiliary variable that activates the first segment of the piecewise function if $\sum_{r\in\reqs}{y_{r,a,s}}<\tilde{y}_{a,s}$, or the second segment otherwise. Note that \eqref{eq:t_inf_piecewise} is quadratic due to the products between $\nu_{a,s}$ and $y_{r',a,s}$. However, these products can be linearized by adding auxiliary variables $\tau_{r',a,s}\in\{0,1\}$ such that $\tau_{r',a,s}\leq\nu_{a,s}$ and $\tau_{r',a,s}\leq y_{r',a,s}$. By combining \eqref{eq:t_inf_piecewise} and its linearization into \eqref{eq:t_inf_con}, we obtain a quadratic constraint due to the product with $\pi_{a,s}$. 





\subsection{Objective function design}\label{sec:objective}


Energy minimization is one of the major drivers of Open \ran, which can scale cloud compute on-the-fly to only activate the resources necessary for service delivery.
To meet these expectations, we consider the total energy cost of activating servers and instantiating \oran applications:
\begin{align}
    E_s(x_s,\mathbf{y}_s) = x_s \cdot E^{\mathrm{base}}_s + \sum_{a\in\apps} \sum_{r\in\reqs} y_{r,a,s} e_{a,s}, \label{eq:energ_server}
\end{align}
\noindent
where $\mathbf{y}_s = (y_{r,a,s})_{(r,a)\in\reqs\times\apps}$, $E^{\mathrm{base}}_s$ represents the fixed amount of energy consumed by server $s$ when turned on (i.e., with at least one application deployed), and $e_{a,s}$ models the energy for an application of type $a$. \eqref{eq:energ_server} is based on experimental evidence showing that energy consumption scales linearly with the server load~\cite{fan2007power}, represented here by the number of applications on the server (last term in \eqref{eq:energ_server}). 
Moreover, $E_s(x_s,\mathbf{y}_s)=0$ when $x_s=0$, and~\eqref{eq:con2} forces all $y_{r,a,s}=0$ to ensure applications selection prioritizes already active servers.

In general, infrastructure owners 
aim at maximizing their profit by minimizing the energy consumed to deliver the most valuable services. 
We formulate such energy-aware profit maximization problem with the following objective function:
\begin{align}
    U(\mathbf{x},\mathbf{y},\mathbf{z}) = \sum_{r\in\reqs} \rho_r z_r - \sigma \sum_{s\in\servers} E_s(x_s,\mathbf{y}_s), \label{eq:ut:energy}
\end{align}
\noindent
where $\rho_r$ represents the monetary payment that the tenant is willing to pay to have their \oran applications deployed on the infrastructure, $\sigma$ is the cost of energy expressed in monetary units per Joule, and $E_s(\cdot)$ is defined in~\eqref{eq:energ_server}. 



\subsection{Computational Complexity}
\label{sec:complexity}

\begin{theorem}\label{th:np_hard}
Problem \eqref{eq:problem} is NP-hard.
\end{theorem}
\noindent
\begin{IEEEproof}
The proof is based on reducing the problem to the quadratically-constrained knapsack problem (QCKP), which is known to be NP-Hard~\cite{klimm2022packing}. Let us consider the case with $S=1$, $\sum_{a\in\apps}n_{r,a}=1$ for all $r\in\reqs$ (i.e., one application per request). Let us assume $\delta_{r,a,s}=1$ for all $(r,a,s)\in\reqs\times\apps\times\servers$, and $L^\mathrm{MAX}_{r,a} = L$ for all $(r,a)\in\reqs\times\apps$, with $L$ a small enough constant that prevents the use of the only server to accommodate all requests. Since each request is associated to one application only, let $\lambda_r = \lambda_a(1)$, where $a$ is the only type of application requested. Recall that the latency function $l_s(\cdot)$ in~\eqref{eq:con3} is an increasing function in the number of requests hosted in each server, and each allocated request contributes with a factor $\lambda_r$ to the total inference time. Problem \eqref{eq:problem} corresponds to an instance of the QCKP with one knapsack (the server) with capacity $L$ (the inference time) and $R$ objects (the requests) of value $\rho_r$ and size $\lambda_r$, with a total value (monetary reward minus the cost) maximization goal. This problem is NP-hard~\cite{klimm2022packing} and we have built a polynomial-time reduction of the QCKP to an
instance of Problem \eqref{eq:problem}. Thus, Problem~\eqref{eq:problem} is NP-hard by reduction unless $\mbox{P}=\mbox{NP}$. 
\end{IEEEproof}


Despite its NP-hardness, Problem \eqref{eq:problem} can be solved optimally via well-established optimization frameworks such as branch-and-bound~\cite{klimm2022packing}. In Sec.~\ref{section:numerical-evaluation}, we show that an optimal solution only requires a few seconds even for large instances of the problem with thousands of O-RAN applications, which is satisfactory and well within the non-real-time requirement of lifecycle management of O-RAN applications~\cite{polese2022understanding}, and we also consider an approximation algorithm that offers lower complexity with slightly lower performance in terms of optimality.

\section{Performance Evaluation}
\label{section:numerical-evaluation}%
We numerically evaluate \name in MATLAB where we solve Problem~\eqref{eq:problem} in Gurobi on a server with an Intel Core i9-9980HK CPU with $16$ cores and $64$\:GB of RAM. For all simulations, we plot results averaged over 50 experiments.

We consider the 3 types of xApps in Table~\ref{table:piecewise} and a conservative fit for \eqref{eq:con3}. The idle energy is $E^{\mathrm{base}}_s=360$\:J (Dell PowerEdge R750) and $e_{a,s}=\{8.77,16,22\}$~J for CNN, LSTM and DRL models by combining the inference/s time from Fig.~\ref{fig:scatter-plot-single-xapps} and the energy consumption per inference in \cite{baskin2018streaming}. The energy cost is $\sigma=0.165$\:\$/kWh (current average in the U.S.). 

We consider three possible \textit{inference time profiles} such that $L_{r,a}\in\{0.2,1,10\}$s and consider the case where $L^\mathrm{MAX}_{r,a}=L_{r,a}$ for all $(r,a)\in\reqs\times\apps$. We do not distinguish between dApps, xApps, and rApps, prioritizing the desired inference time required by each tenant. We refer to the above inference time profiles as \gls{rt}, \nearrt and \nonrt, respectively. For each request $r$, we set $n_{r,a'}=n_{r,a''}$ for any $(a',a'')$ and randomly select one inference time demand from the set defined above with probability $10\%$, $60\%$, $30\%$ for \gls{rt}, \nearrt and \nonrt, respectively. In the following, we present results as a function of the \textit{total number of instances} requested by all tenants which is defined as $I \! = \! \sum_{r\in\reqs}\sum_{a\in\apps} n_{r,a}$. 
We consider homogeneous requests with same monetary value $\rho_r \! = \! 2$\$ and same total number $\sum_{a\in\apps} n_{r,a}$ of application instances requested.
The number of requests is $R \! = \! 5$ and we vary $\sum_{a\in\apps} n_{r,a}$ to emulate very small or very large numbers of \ai models for the control of a certain O-RAN deployment.

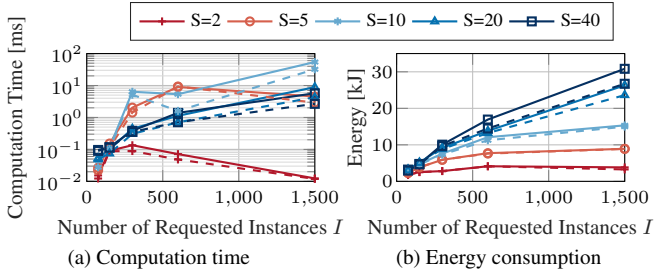
\begin{figure}[t!]
\setlength\abovecaptionskip{1pt}
\ifexttikz
    \tikzsetnextfilename{complexity}
\fi
\begin{subfigure}[t]{0.48\columnwidth}
    \setlength\abovecaptionskip{0pt}
    \centering
    \setlength\fwidth{0.75\columnwidth}
    \setlength\fheight{0.4\columnwidth}
%
%
\begin{tikzpicture}
\pgfplotsset{every tick label/.append style={font=\scriptsize}}

\begin{axis}[%
width=0.951\fwidth,
height=\fheight,
at={(0\fwidth,0\fheight)},
scale only axis,
xmin=0,
xmax=1500,
xlabel={Number of Requested Instances $I$},
ymin=0.01,
ymax=100,
ymode=log,
%
ytick distance=10,
minor grid style={very thin,gray!30},
grid=both,
ylabel={Computation Time [ms]},
axis background/.style={fill=white},
xmajorgrids,
ymajorgrids,
legend style={legend cell align=left, align=left, draw=white!15!black,
anchor=south, font=\scriptsize, align=left, at={(1.25,1.1)}},
legend columns=5,
xlabel style={font=\footnotesize\color{white!15!black}},
ylabel style={font=\footnotesize\color{white!15!black}},
yticklabel shift=-2pt,
ylabel shift=-3pt,
xlabel shift=-2pt,
]
\addplot [color=RdBu-B, thick, mark=+, mark size=1.5pt, mark options={solid, RdBu-B}]
  table[row sep=crcr]{%
75	0.0153038740158081\\
150	0.0905955696105957\\
300	0.135213646888733\\
600	0.0713622856140137\\
1500	0.0123763036727905\\
};
\addlegendentry{S=2}

\addplot [color=RdBu-D, thick, mark=o, mark size=1.5pt, mark options={solid, RdBu-D}]
  table[row sep=crcr]{%
75	0.0219468259811401\\
150	0.152811374664307\\
300	2.03367880821228\\
600	9.03019049167633\\
1500	4.53102527618408\\
};
\addlegendentry{S=5}

\addplot [color=RdBu-K, thick, mark=asterisk, mark size=1.5pt, mark options={solid, RdBu-K}]
  table[row sep=crcr]{%
75	0.0356357002258301\\
150	0.115338973999023\\
300	6.3731036901474\\
600	5.34574652194977\\
1500	54.1577863121033\\
};
\addlegendentry{S=10}

\addplot [color=RdBu-M, thick, mark=triangle, mark size=1.5pt, mark options={solid, RdBu-M}]
  table[row sep=crcr]{%
75	0.053619909286499\\
150	0.0992148494720459\\
300	0.443618898391724\\
600	1.12724242210388\\
1500	8.86500762462616\\
};
\addlegendentry{S=20}

\addplot [color=RdBu-O, thick, mark=square, mark size=1.5pt, mark options={solid, RdBu-O}]
  table[row sep=crcr]{%
75	0.0961979055404663\\
150	0.118197693824768\\
300	0.383656687736511\\
600	1.39125665187836\\
1500	5.71087140083313\\
};
\addlegendentry{S=40}

\addplot [color=RdBu-B, thick, dashed, mark=+, mark size=1.5pt, mark options={solid, RdBu-B}, forget plot]
  table[row sep=crcr]{%
75	0.0122816753387451\\
150	0.087628493309021\\
300	0.0887370204925537\\
600	0.0484190559387207\\
1500	0.0119062662124634\\
};
\addplot [color=RdBu-D, thick, dashed, mark=o, mark size=1.5pt, mark options={solid, RdBu-D}, forget plot]
  table[row sep=crcr]{%
75	0.0264095973968506\\
150	0.113556656837463\\
300	1.42118340969086\\
600	9.31167418479919\\
1500	2.91630983829498\\
};
\addplot [color=RdBu-K, thick, dashed, mark=asterisk, mark size=1.5pt, mark options={solid, RdBu-K}, forget plot]
  table[row sep=crcr]{%
75	0.0276617240905762\\
150	0.104038515090942\\
300	5.11089997768402\\
600	1.60904620170593\\
1500	32.1617273139954\\
};
\addplot [color=RdBu-M, thick, dashed, mark=triangle, mark size=1.5pt, mark options={solid, RdBu-M}, forget plot]
  table[row sep=crcr]{%
75	0.0497565126419067\\
150	0.0746009302139282\\
300	0.306437182426453\\
600	0.70404869556427\\
1500	4.29461763858795\\
};
\addplot [color=RdBu-O, thick, dashed, mark=square, mark size=1.5pt, mark options={solid, RdBu-O}, forget plot]
  table[row sep=crcr]{%
75	0.0898730039596558\\
150	0.113191027641296\\
300	0.352980284690857\\
600	0.709004173278809\\
1500	2.66636281013489\\
};

\end{axis}

\end{tikzpicture}%
    \caption{Computation time}
    \label{fig:complexity}
\end{subfigure}\hfill
\ifexttikz
    \tikzsetnextfilename{energy}
\fi
\begin{subfigure}[t]{0.48\columnwidth}
    \setlength\abovecaptionskip{0pt}
    \centering
    \setlength\fwidth{0.75\columnwidth}
    \setlength\fheight{0.4\columnwidth}
%
%
\begin{tikzpicture}
\pgfplotsset{every tick label/.append style={font=\scriptsize}}

\begin{axis}[%
width=0.951\fwidth,
height=\fheight,
at={(0\fwidth,0\fheight)},
scale only axis,
xmin=0,
xmax=1500,
xlabel={Number of Requested Instances $I$},
ymin=0,
ymax=35,
ylabel={Energy [kJ]},
axis background/.style={fill=white},
xmajorgrids,
ymajorgrids,
legend style={legend cell align=left, align=left, draw=white!15!black,
anchor=south, font=\scriptsize, align=left, at={(1.25,1.1)}},
legend columns=5,
xlabel style={font=\footnotesize\color{white!15!black}},
ylabel style={font=\footnotesize\color{white!15!black}},
yticklabel shift=-2pt,
ylabel shift=-5pt,
xlabel shift=-2pt,
]
\addplot [color=RdBu-B, thick, mark=+, mark size=1.5pt, mark options={solid, RdBu-B}]
  table[row sep=crcr]{%
75	1.79570999998617\\
150	2.55338399999684\\
300	2.80809599999737\\
600	4.13488799999671\\
1500	3.82811999999988\\
};
\addlegendentry{S=2}

\addplot [color=RdBu-D, thick, mark=o, mark size=1.5pt, mark options={solid, RdBu-D}]
  table[row sep=crcr]{%
75	2.27805000000852\\
150	3.84114599999944\\
300	5.89699200000201\\
600	7.7217599999965\\
1500	8.91587999999921\\
};
\addlegendentry{S=5}

\addplot [color=RdBu-K, thick, mark=asterisk, mark size=1.5pt, mark options={solid, RdBu-K}]
  table[row sep=crcr]{%
75	2.75324999999317\\
150	4.71449999999539\\
300	7.60173599999829\\
600	12.0889440000031\\
1500	15.3353999999949\\
};
\addlegendentry{S=10}

\addplot [color=RdBu-M, thick, mark=triangle, mark size=1.5pt, mark options={solid, RdBu-M}]
  table[row sep=crcr]{%
75	3.27885000000172\\
150	5.14649999999842\\
300	9.17699999999684\\
600	13.829808000001\\
1500	26.2959599999966\\
};
\addlegendentry{S=20}

\addplot [color=RdBu-O, thick, mark=square, mark size=1.5pt, mark options={solid, RdBu-O}]
  table[row sep=crcr]{%
75	3.22844999999719\\
150	4.77210000000189\\
300	10.0554000000014\\
600	16.9271039999961\\
1500	30.8238599999961\\
};
\addlegendentry{S=40}

\addplot [color=RdBu-B, thick, dashed, mark=+, mark size=1.5pt, mark options={solid, RdBu-B}, forget plot]
  table[row sep=crcr]{%
75	1.79570999998423\\
150	2.51596800000003\\
300	2.81529599999702\\
600	4.12048799999635\\
1500	3.26688000000059\\
};
\addplot [color=RdBu-D, thick, dashed, mark=o, mark size=1.5pt, mark options={solid, RdBu-D}, forget plot]
  table[row sep=crcr]{%
75	2.32125000001185\\
150	3.81739199999474\\
300	5.92579200000255\\
600	7.60231199999924\\
1500	8.87988000000038\\
};
\addplot [color=RdBu-K, thick, dashed, mark=asterisk, mark size=1.5pt, mark options={solid, RdBu-K}, forget plot]
  table[row sep=crcr]{%
75	2.77484999999444\\
150	4.61659200000239\\
300	7.35268799999938\\
600	11.2570559999988\\
1500	14.98278\\
};
\addplot [color=RdBu-M, thick, dashed, mark=triangle, mark size=1.5pt, mark options={solid, RdBu-M}, forget plot]
  table[row sep=crcr]{%
75	2.99804999999987\\
150	5.06730000000072\\
300	8.85736799999845\\
600	13.2323279999989\\
1500	23.6190600000013\\
};
\addplot [color=RdBu-O, thick, dashed, mark=square, mark size=1.5pt, mark options={solid, RdBu-O}, forget plot]
  table[row sep=crcr]{%
75	2.95484999999887\\
150	4.65689999999974\\
300	9.66518399999733\\
600	14.4808799999993\\
1500	26.7289799999976\\
};

\legend{}  

\end{axis}

\end{tikzpicture}%
    \caption{Energy consumption}
    \label{fig:energy}
\end{subfigure}
\setlength\belowcaptionskip{-.3cm}
\caption{Scalability and energy analysis for varying number of servers ($S$) and number of instances ($I$). Solid lines: optimal solution; dashed: early stopping.}
\label{fig:complexity_energy}
\end{figure}

\begin{figure}[t]
    \setlength\abovecaptionskip{0pt}
    \centering
    \setlength\fwidth{0.95\columnwidth}
    \setlength\fheight{0.15\columnwidth}
%
%
\begin{tikzpicture}
\pgfplotsset{every tick label/.append style={font=\scriptsize}}

\begin{axis}[%
width=0.247\fwidth,
height=\fheight,
at={(0\fwidth,0\fheight)},
scale only axis,
bar shift auto,
xmin=0.511111111111111,
xmax=5.48888888888889,
xtick={1,2,3,4,5},
xticklabels={{75},{150},{300},{600},{1500}},
xticklabel style={rotate=90},
xlabel style={font=\footnotesize\color{white!15!black}},
ymin=0,
ymax=0.9,
ylabel style={font=\footnotesize\color{white!15!black}},
ylabel={Application Presence},
axis background/.style={fill=white},
title style={font=\bfseries},
xmajorgrids,
ymajorgrids,
yticklabel shift=-2pt,
ylabel shift=-3pt,
xlabel shift=-2pt,
ybar=0pt,
xtick align=inside
]
\addplot[ybar, bar width=0.25, fill=Blues-F, draw=Blues-F, area legend] table[row sep=crcr] {%
1	0\\
2	0\\
3	0\\
4	0\\
5	nan\\
};
\addplot[forget plot, color=white!15!black] table[row sep=crcr] {%
0.511111111111111	0\\
5.48888888888889	0\\
};
\addplot[ybar, bar width=0.25, fill=Blues-I, draw=Blues-I, area legend] table[row sep=crcr] {%
1	0.653009708917988\\
2	0.741933982282875\\
3	0.379485677798335\\
4	0.14\\
5	nan\\
};
\addplot[forget plot, color=white!15!black] table[row sep=crcr] {%
0.511111111111111	0\\
5.48888888888889	0\\
};
\addplot[ybar, bar width=0.25, fill=Blues-L, draw=Blues-L, area legend] table[row sep=crcr] {%
1	0.346990291082012\\
2	0.258066017717125\\
3	0.620514322201665\\
4	0.86\\
5	nan\\
};
\addplot[forget plot, color=white!15!black] table[row sep=crcr] {%
0.511111111111111	0\\
5.48888888888889	0\\
};

\node[align=left,font=\scriptsize\linespread{0.9}\selectfont,yshift=1pt,fill=white,draw=black,inner sep=1.2pt] (off) at (axis cs: 3,0.75) {S=2};

\end{axis}

\begin{axis}[%
width=0.247\fwidth,
height=\fheight,
at={(0.346\fwidth,0\fheight)},
scale only axis,
bar shift auto,
xmin=0.511111111111111,
xmax=5.48888888888889,
xtick={1,2,3,4,5},
xticklabels={{75},{150},{300},{600},{1500}},
xticklabel style={rotate=90},
xlabel style={font=\footnotesize\color{white!15!black}},
xlabel={Number of Requested Instances $I$},
ymin=0,
ymax=0.9,
ylabel style={font=\footnotesize\color{white!15!black}},
axis background/.style={fill=white},
title style={font=\bfseries},
xmajorgrids,
ymajorgrids,
yticklabel shift=-2pt,
ylabel shift=-3pt,
xlabel shift=-2pt,
ybar=0pt,
xtick align=inside
]
\addplot[ybar, bar width=0.25, fill=Blues-F, draw=Blues-F, area legend] table[row sep=crcr] {%
1	0.217918531176426\\
2	0.281067678548539\\
3	0.222593760041128\\
4	0\\
5	0\\
};
\addplot[forget plot, color=white!15!black] table[row sep=crcr] {%
0.511111111111111	0\\
5.48888888888889	0\\
};
\addplot[ybar, bar width=0.25, fill=Blues-I, draw=Blues-I, area legend] table[row sep=crcr] {%
1	0.52352531292619\\
2	0.47710483219827\\
3	0.512137618326253\\
4	0.806251549161891\\
5	0.693056138311545\\
};
\addplot[forget plot, color=white!15!black] table[row sep=crcr] {%
0.511111111111111	0\\
5.48888888888889	0\\
};
\addplot[ybar, bar width=0.25, fill=Blues-L, draw=Blues-L, area legend] table[row sep=crcr] {%
1	0.258556155897384\\
2	0.239327489253191\\
3	0.265268621632618\\
4	0.193748450838109\\
5	0.306943861688455\\
};
\addplot[forget plot, color=white!15!black] table[row sep=crcr] {%
0.511111111111111	0\\
5.48888888888889	0\\
};

\node[align=left,font=\scriptsize\linespread{0.9}\selectfont,yshift=1pt,fill=white,draw=black,inner sep=1.2pt] (off) at (axis cs: 3,0.75) {S=10};

\end{axis}

\begin{axis}[%
width=0.247\fwidth,
height=\fheight,
at={(0.692\fwidth,0\fheight)},
scale only axis,
bar shift auto,
xmin=0.511111111111111,
xmax=5.48888888888889,
xtick={1,2,3,4,5},
xticklabels={{75},{150},{300},{600},{1500}},
xticklabel style={rotate=90},
xlabel style={font=\footnotesize\color{white!15!black}},
ymin=0,
ymax=0.9,
ylabel style={font=\footnotesize\color{white!15!black}},
axis background/.style={fill=white},
title style={font=\bfseries},
xmajorgrids,
ymajorgrids,
yticklabel shift=-2pt,
ylabel shift=-3pt,
xlabel shift=-2pt,
legend columns=3,
legend style={legend cell align=left, align=left, draw=white!15!black,
anchor=south, font=\scriptsize, align=left, at={(-1,1.05)}},
ybar=0pt,
xtick align=inside
]
\addplot[ybar, bar width=0.25, fill=Blues-F, draw=Blues-F, area legend] table[row sep=crcr] {%
1	0.206485215065865\\
2	0.177921214664249\\
3	0.195306359769494\\
4	0.210182834836006\\
5	0.0276154053224229\\
};
\addplot[forget plot, color=white!15!black] table[row sep=crcr] {%
0.511111111111111	0\\
5.48888888888889	0\\
};
\addlegendentry{RT}  

\addplot[ybar, bar width=0.25, fill=Blues-I, draw=Blues-I, area legend] table[row sep=crcr] {%
1	0.526977891787721\\
2	0.564673470740358\\
3	0.514179830834248\\
4	0.621506650568963\\
5	0.806985045199537\\
};
\addplot[forget plot, color=white!15!black] table[row sep=crcr] {%
0.511111111111111	0\\
5.48888888888889	0\\
};
\addlegendentry{Near-RT}  

\addplot[ybar, bar width=0.25, fill=Blues-L, draw=Blues-L, area legend] table[row sep=crcr] {%
1	0.266536893146413\\
2	0.257405314595393\\
3	0.259218409157309\\
4	0.159889461963452\\
5	0.157028815490195\\
};
\addplot[forget plot, color=white!15!black] table[row sep=crcr] {%
0.511111111111111	0\\
5.48888888888889	0\\
};
\addlegendentry{Non-RT}  

\node[align=left,font=\scriptsize\linespread{0.9}\selectfont,yshift=1pt,fill=white,draw=black,inner sep=1.2pt] (off) at (axis cs: 3,0.75) {S=40};

\end{axis}

\end{tikzpicture}%
    \setlength\abovecaptionskip{-.3cm}
    \setlength\belowcaptionskip{-.3cm}
    \caption{Probability of a server to host a request with a certain inference time profile for varying number of servers ($S$) and number of instances ($I$).}
    \label{fig:presence}
\end{figure}
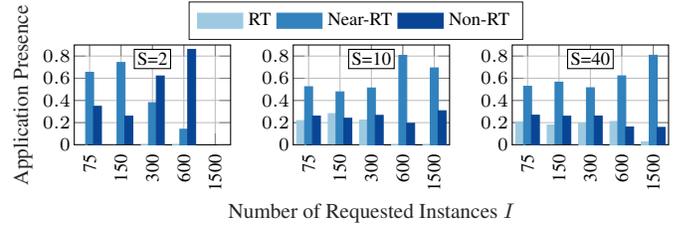

First, we analyze the \textit{complexity} of solving Problem~\eqref{eq:problem} optimally (solid lines in the figures). Fig.~\ref{fig:complexity} shows the computation time as a function of $I$ for different number of servers $S$. 
Intuitively, the complexity grows with the number of \ai models ($I$) to deploy up to a threshold $I^*$, where the trend reverses. As we show next (Fig.~\ref{fig:presence}), this happens because for large $I$ the optimization engine neglects requests with \gls{rt} and \nearrt inference profiles, prioritizing \nonrt requests which can be satisfied in larger numbers. Indeed, the cost for accommodating \gls{rt} and \nearrt requests is too high (they force a limit on the inference time for the entire server) as it prevents the admission of \nonrt requests. Thus, the algorithm discards their branches, converging faster to an optimal solution. We also compare against an approximation approach (dashed lines) where we perform early stopping on the branch-and-bound procedure when all reduced costs of the underlying dual problem are less than $0.01$. As expected, early stopping produces sub-optimal solutions in less time, with a $2.16\times$ gain when $S=40$.

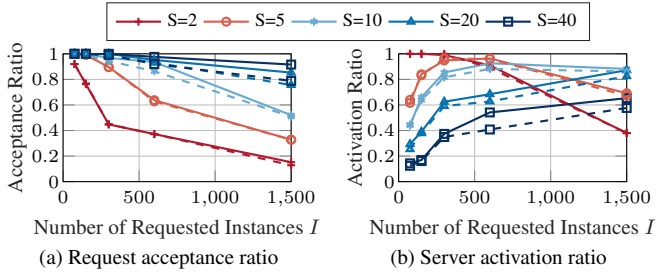
\begin{figure}[t]
\setlength\abovecaptionskip{1pt}
\ifexttikz
    \tikzsetnextfilename{acceptance}
\fi
\begin{subfigure}[t]{0.48\columnwidth}
    \setlength\abovecaptionskip{0pt}
    \centering
    \setlength\fwidth{0.75\columnwidth}
    \setlength\fheight{0.4\columnwidth}
%
%
\begin{tikzpicture}
\pgfplotsset{every tick label/.append style={font=\scriptsize}}

\begin{axis}[%
width=0.951\fwidth,
height=\fheight,
at={(0\fwidth,0\fheight)},
scale only axis,
xmin=0,
xmax=1500,
xlabel={Number of Requested Instances $I$},
ymin=0,
ymax=1,
ylabel={Acceptance Ratio},
axis background/.style={fill=white},
xmajorgrids,
ymajorgrids,
legend style={legend cell align=left, align=left, draw=white!15!black,
anchor=south, font=\scriptsize, align=left, at={(1.25,1.1)}},
legend columns=5,
xlabel style={font=\footnotesize\color{white!15!black}},
ylabel style={font=\footnotesize\color{white!15!black}},
yticklabel shift=-2pt,
ylabel shift=-5pt,
xlabel shift=-2pt,
]
\addplot [color=RdBu-B, thick, mark=+, mark size=1.5pt, mark options={solid, RdBu-B}]
  table[row sep=crcr]{%
75	0.92\\
150	0.762666666666667\\
300	0.448\\
600	0.372\\
1500	0.152\\
};
\addlegendentry{S=2}

\addplot [color=RdBu-D, thick, mark=o, mark size=1.5pt, mark options={solid, RdBu-D}]
  table[row sep=crcr]{%
75	1\\
150	0.993333333333333\\
300	0.896\\
600	0.637333333333333\\
1500	0.328\\
};
\addlegendentry{S=5}

\addplot [color=RdBu-K, thick, mark=asterisk, mark size=1.5pt, mark options={solid, RdBu-K}]
  table[row sep=crcr]{%
75	1\\
150	1\\
300	0.968\\
600	0.936\\
1500	0.52\\
};
\addlegendentry{S=10}

\addplot [color=RdBu-M, thick, mark=triangle, mark size=1.5pt, mark options={solid, RdBu-M}]
  table[row sep=crcr]{%
75	0.996\\
150	1\\
300	1\\
600	0.952\\
1500	0.853333333333333\\
};
\addlegendentry{S=20}

\addplot [color=RdBu-O, thick, mark=square, mark size=1.5pt, mark options={solid, RdBu-O}]
  table[row sep=crcr]{%
75	1\\
150	1\\
300	1\\
600	0.976\\
1500	0.916\\
};
\addlegendentry{S=40}

\addplot [color=RdBu-B, thick, dashed, mark=+, mark size=1.5pt, mark options={solid, RdBu-B}, forget plot]
  table[row sep=crcr]{%
75	0.92\\
150	0.768\\
300	0.448\\
600	0.372\\
1500	0.128\\
};
\addplot [color=RdBu-D, thick, dashed, mark=o, mark size=1.5pt, mark options={solid, RdBu-D}, forget plot]
  table[row sep=crcr]{%
75	1\\
150	0.992\\
300	0.896\\
600	0.628\\
1500	0.328\\
};
\addplot [color=RdBu-K, thick, dashed, mark=asterisk, mark size=1.5pt, mark options={solid, RdBu-K}, forget plot]
  table[row sep=crcr]{%
75	1\\
150	0.992\\
300	0.944\\
600	0.864\\
1500	0.508\\
};
\addplot [color=RdBu-M, thick, dashed, mark=triangle, mark size=1.5pt, mark options={solid, RdBu-M}, forget plot]
  table[row sep=crcr]{%
75	1\\
150	1\\
300	0.984\\
600	0.932\\
1500	0.756\\
};
\addplot [color=RdBu-O, thick, dashed, mark=square, mark size=1.5pt, mark options={solid, RdBu-O}, forget plot]
  table[row sep=crcr]{%
75	1\\
150	1\\
300	0.992\\
600	0.92\\
1500	0.788\\
};

\end{axis}

\end{tikzpicture}%
    \caption{Request acceptance ratio}
    \label{fig:acceptance}
\end{subfigure}\hfill
\ifexttikz
    \tikzsetnextfilename{activation}
\fi
\begin{subfigure}[t]{0.48\columnwidth}
    \setlength\abovecaptionskip{0pt}
    \centering
    \setlength\fwidth{0.75\columnwidth}
    \setlength\fheight{0.4\columnwidth}
%
%
\begin{tikzpicture}
\pgfplotsset{every tick label/.append style={font=\scriptsize}}

\begin{axis}[%
width=0.951\fwidth,
height=\fheight,
at={(0\fwidth,0\fheight)},
scale only axis,
xmin=0,
xmax=1500,
xlabel={Number of Requested Instances $I$},
ymin=0,
ymax=1,
ylabel={Activation Ratio},
axis background/.style={fill=white},
xmajorgrids,
ymajorgrids,
legend style={legend cell align=left, align=left, draw=white!15!black,
anchor=north west, font=\scriptsize, align=left, at={(0.02,0.98)}},
legend columns=1,
xlabel style={font=\footnotesize\color{white!15!black}},
ylabel style={font=\footnotesize\color{white!15!black}},
yticklabel shift=-2pt,
ylabel shift=-5pt,
xlabel shift=-2pt,
]
\addplot [color=RdBu-B, thick, mark=+, mark size=1.5pt, mark options={solid, RdBu-B}]
  table[row sep=crcr]{%
75	1\\
150	1\\
300	0.99\\
600	0.91\\
1500	0.38\\
};
\addlegendentry{S=2}

\addplot [color=RdBu-D, thick, mark=o, mark size=1.5pt, mark options={solid, RdBu-D}]
  table[row sep=crcr]{%
75	0.616\\
150	0.84\\
300	0.948\\
600	0.964\\
1500	0.692\\
};
\addlegendentry{S=5}

\addplot [color=RdBu-K, thick, mark=asterisk, mark size=1.5pt, mark options={solid, RdBu-K}]
  table[row sep=crcr]{%
75	0.44\\
150	0.66\\
300	0.854\\
600	0.926\\
1500	0.882\\
};
\addlegendentry{S=10}

\addplot [color=RdBu-M, thick, mark=triangle, mark size=1.5pt, mark options={solid, RdBu-M}]
  table[row sep=crcr]{%
75	0.293\\
150	0.39\\
300	0.625\\
600	0.684\\
1500	0.872\\
};
\addlegendentry{S=20}

\addplot [color=RdBu-O, thick, mark=square, mark size=1.5pt, mark options={solid, RdBu-O}]
  table[row sep=crcr]{%
75	0.143\\
150	0.169\\
300	0.3735\\
600	0.5415\\
1500	0.653\\
};
\addlegendentry{S=40}

\addplot [color=RdBu-B, thick, dashed, mark=+, mark size=1.5pt, mark options={solid, RdBu-B}, forget plot]
  table[row sep=crcr]{%
75	1\\
150	1\\
300	1\\
600	0.89\\
1500	0.38\\
};
\addplot [color=RdBu-D, thick, dashed, mark=o, mark size=1.5pt, mark options={solid, RdBu-D}, forget plot]
  table[row sep=crcr]{%
75	0.64\\
150	0.832\\
300	0.964\\
600	0.96\\
1500	0.672\\
};
\addplot [color=RdBu-K, thick, dashed, mark=asterisk, mark size=1.5pt, mark options={solid, RdBu-K}, forget plot]
  table[row sep=crcr]{%
75	0.446\\
150	0.638\\
300	0.816\\
600	0.882\\
1500	0.862\\
};
\addplot [color=RdBu-M, thick, dashed, mark=triangle, mark size=1.5pt, mark options={solid, RdBu-M}, forget plot]
  table[row sep=crcr]{%
75	0.254\\
150	0.379\\
300	0.591\\
600	0.627\\
1500	0.825\\
};
\addplot [color=RdBu-O, thick, dashed, mark=square, mark size=1.5pt, mark options={solid, RdBu-O}, forget plot]
  table[row sep=crcr]{%
75	0.124\\
150	0.161\\
300	0.349\\
600	0.408\\
1500	0.5765\\
};

\legend{}  

\end{axis}

\end{tikzpicture}%
    \caption{Server activation ratio}
    \label{fig:activation}
\end{subfigure}
\setlength\belowcaptionskip{-.3cm}
\caption{Acceptance and activation ratios for varying number of servers ($S$) and number of instances ($I$). Solid lines: optimal solution; dashed: early stopping.}
\label{fig:acceptance_activation}
\end{figure}

Fig.~\ref{fig:energy} shows that the total \textit{energy consumption} always increases with $I$, with a plateau when no more requests can be admitted.
Early stopping computes solutions that consume less energy than the optimal. This is a consequence of its lower \emph{acceptance ratio} (i.e., the ratio between the number of \ai models actually instantiated and $I$, Fig.~\ref{fig:acceptance}) and lower \textit{activation ratio} (i.e., the percentage of servers that host at least one application, Fig.~\ref{fig:activation}). 
The optimal solution satisfies more than 90\% of requests, with 65\% servers activated when $S=40$.

Fig.~\ref{fig:acceptance} shows that the acceptance ratio decreases with $I$ and increases with larger number of available servers $S$. Differently, the activation ratio trend (Fig.~\ref{fig:activation}) is similar to that of the complexity (Fig.~\ref{fig:complexity}). Indeed, when the number of servers $S$ in the cluster is small, the activation ratio decreases with $I$, as it becomes impossible to allocate even a single \gls{rt} or \nearrt without violating \eqref{eq:con3} with high probability.


\ifexttikz
    \tikzsetnextfilename{presence}
\fi

Fig.~\ref{fig:presence} shows the \textit{application presence} probability, i.e., the probability that requests with diverse inference time profiles are admitted by \name. When $S=2$, \gls{rt} requests are completely neglected, as they limit the number of admissible \ai models.\footnote{Constraint \eqref{eq:con3} forces a server to satisfy the latency requirement of the most demanding application being hosted in the server.} Indeed, we see that with more servers it is possible to admit more \gls{rt} and \nearrt requests. \textit{These results clearly show that scaling \ai solutions in \oran systems is not a resource-constrained problem, but a time-constrained one in that requirements on the inference time strongly affect how many dApps, xApps and rApps can coexist on the same server.}

\begin{figure}[t]
    \setlength\abovecaptionskip{0pt}
    \centering
    \setlength\fwidth{0.95\columnwidth}
    \setlength\fheight{0.15\columnwidth}
%
%
\begin{tikzpicture}
\pgfplotsset{every tick label/.append style={font=\scriptsize}}

\begin{axis}[%
width=0.247\fwidth,
height=\fheight,
at={(0\fwidth,0\fheight)},
scale only axis,
bar shift auto,
xmin=0.511111111111111,
xmax=5.48888888888889,
xtick={1,2,3,4,5},
xticklabels={{75},{150},{300},{600},{1500}},
xticklabel style={rotate=90},
xlabel style={font=\footnotesize\color{white!15!black}},
ymin=0,
ymax=0.9,
ylabel style={font=\footnotesize\color{white!15!black}},
ylabel={Coexistence Factor},
axis background/.style={fill=white},
title style={font=\bfseries},
xmajorgrids,
ymajorgrids,
yticklabel shift=-2pt,
ylabel shift=-3pt,
xlabel shift=-2pt,
ybar=0pt,
xtick align=inside
]
\addplot[ybar, bar width=0.25, fill=Blues-F, draw=Blues-F, area legend] table[row sep=crcr] {%
1	0\\
2	0\\
3	0\\
4	0\\
5	0\\
};
\addplot[forget plot, color=white!15!black] table[row sep=crcr] {%
0.511111111111111	0\\
5.48888888888889	0\\
};
\addplot[ybar, bar width=0.25, fill=Blues-I, draw=Blues-I, area legend] table[row sep=crcr] {%
1	0.88\\
2	0.64\\
3	0.2\\
4	0\\
5	0\\
};
\addplot[forget plot, color=white!15!black] table[row sep=crcr] {%
0.511111111111111	0\\
5.48888888888889	0\\
};
\addplot[ybar, bar width=0.25, fill=Blues-L, draw=Blues-L, area legend] table[row sep=crcr] {%
1	0.88\\
2	0.61\\
3	0.2\\
4	0\\
5	0\\
};
\addplot[forget plot, color=white!15!black] table[row sep=crcr] {%
0.511111111111111	0\\
5.48888888888889	0\\
};

\node[align=left,font=\scriptsize\linespread{0.9}\selectfont,yshift=1pt,fill=white,draw=black,inner sep=1.2pt] (off) at (axis cs: 3.2,0.75) {S=2};

\end{axis}

\begin{axis}[%
width=0.247\fwidth,
height=\fheight,
at={(0.346\fwidth,0\fheight)},
scale only axis,
bar shift auto,
xmin=0.511111111111111,
xmax=5.48888888888889,
xtick={1,2,3,4,5},
xticklabels={{75},{150},{300},{600},{1500}},
xticklabel style={rotate=90},
xlabel style={font=\footnotesize\color{white!15!black}},
xlabel={Number of Requested Instances $I$},
ymin=0,
ymax=0.9,
ylabel style={font=\footnotesize\color{white!15!black}},
axis background/.style={fill=white},
title style={font=\bfseries},
xmajorgrids,
ymajorgrids,
yticklabel shift=-2pt,
ylabel shift=-3pt,
xlabel shift=-2pt,
ybar=0pt,
xtick align=inside
]
\addplot[ybar, bar width=0.25, fill=Blues-F, draw=Blues-F, area legend] table[row sep=crcr] {%
1	0.158166666666667\\
2	0.243126984126984\\
3	0.217023809523809\\
4	0\\
5	0\\
};
\addplot[forget plot, color=white!15!black] table[row sep=crcr] {%
0.511111111111111	0\\
5.48888888888889	0\\
};
\addplot[ybar, bar width=0.25, fill=Blues-I, draw=Blues-I, area legend] table[row sep=crcr] {%
1	0.773452380952381\\
2	0.747666666666667\\
3	0.546039682539682\\
4	0.12831746031746\\
5	0.317047619047619\\
};
\addplot[forget plot, color=white!15!black] table[row sep=crcr] {%
0.511111111111111	0\\
5.48888888888889	0\\
};
\addplot[ybar, bar width=0.25, fill=Blues-L, draw=Blues-L, area legend] table[row sep=crcr] {%
1	0.725047619047619\\
2	0.7835\\
3	0.586849206349206\\
4	0.227952380952381\\
5	0.313595238095238\\
};
\addplot[forget plot, color=white!15!black] table[row sep=crcr] {%
0.511111111111111	0\\
5.48888888888889	0\\
};

\node[align=left,font=\scriptsize\linespread{0.9}\selectfont,yshift=1pt,fill=white,draw=black,inner sep=1.2pt] (off) at (axis cs: 3.2,0.75) {S=10};

\end{axis}

\begin{axis}[%
width=0.247\fwidth,
height=\fheight,
at={(0.692\fwidth,0\fheight)},
scale only axis,
bar shift auto,
xmin=0.511111111111111,
xmax=5.48888888888889,
xtick={1,2,3,4,5},
xticklabels={{75},{150},{300},{600},{1500}},
xticklabel style={rotate=90},
xlabel style={font=\footnotesize\color{white!15!black}},
ymin=0,
ymax=0.9,
ylabel style={font=\footnotesize\color{white!15!black}},
axis background/.style={fill=white},
title style={font=\bfseries},
xmajorgrids,
ymajorgrids,
yticklabel shift=-2pt,
ylabel shift=-3pt,
xlabel shift=-2pt,
legend columns=3,
legend style={legend cell align=left, align=left, draw=white!15!black,
anchor=south, font=\scriptsize, align=left, at={(-1,1.05)}},
ybar=0pt,
xtick align=inside
]
\addplot[ybar, bar width=0.25, fill=Blues-F, draw=Blues-F, area legend] table[row sep=crcr] {%
1	0.1785\\
2	0.0676666666666667\\
3	0.14623778998779\\
4	0.158665576882846\\
5	0.0137379310344828\\
};
\addplot[forget plot, color=white!15!black] table[row sep=crcr] {%
0.511111111111111	0\\
5.48888888888889	0\\
};
\addlegendentry{RT}  %

\addplot[ybar, bar width=0.25, fill=Blues-I, draw=Blues-I, area legend] table[row sep=crcr] {%
1	0.846777777777778\\
2	0.660333333333333\\
3	0.494021520624969\\
4	0.249864267890858\\
5	0.116212720013649\\
};
\addplot[forget plot, color=white!15!black] table[row sep=crcr] {%
0.511111111111111	0\\
5.48888888888889	0\\
};
\addlegendentry{Near-RT}  

\addplot[ybar, bar width=0.25, fill=Blues-L, draw=Blues-L, area legend] table[row sep=crcr] {%
1	0.836666666666667\\
2	0.766285714285714\\
3	0.556060690943044\\
4	0.256444494720965\\
5	0.222606496444732\\
};
\addplot[forget plot, color=white!15!black] table[row sep=crcr] {%
0.511111111111111	0\\
5.48888888888889	0\\
};
\addlegendentry{Non-RT}  

\node[align=left,font=\scriptsize\linespread{0.9}\selectfont,yshift=1pt,fill=white,draw=black,inner sep=1.2pt] (off) at (axis cs: 3.2,0.75) {S=40};

\end{axis}

\end{tikzpicture}%
    \caption{Probability that requests with a certain inference time profile can coexist on the same server for varying number of servers ($S$) and instances ($I$).}
    \label{fig:coexistence}
\vspace{-5pt}
\end{figure}
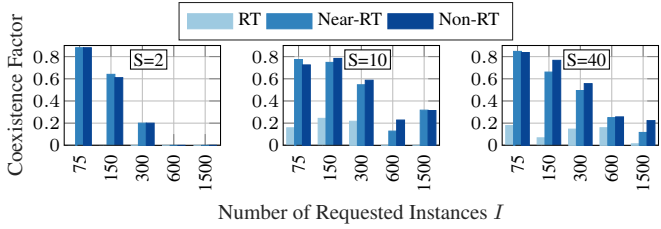

Fig.~\ref{fig:coexistence} shows the probability that requests with diverse inference time profiles coexist on the same server. \gls{rt} requests are less likely to share the same server with other profiles. For $I\leq 300$, both \nearrt and \nonrt requests can coexist with a probability higher than $0.5$, which however drops to approximately $0.2$ when $I$ is large.

\begin{figure}[t]
    \setlength\abovecaptionskip{0pt}
    \centering
    \setlength\fwidth{0.95\columnwidth}
    \setlength\fheight{0.15\columnwidth}
%
%
\begin{tikzpicture}
\pgfplotsset{every tick label/.append style={font=\scriptsize}}

\begin{axis}[%
width=0.167\fwidth,
height=\fheight,
at={(0\fwidth,0\fheight)},
scale only axis,
bar shift auto,
xmin=2.51111111111111,
xmax=5.48888888888889,
xtick={3,4,5},
xticklabels={{300},{600},{1500}},
xticklabel style={rotate=90},
xlabel style={font=\footnotesize\color{white!15!black}},
ymin=0,
ymax=1,
ylabel style={font=\footnotesize\color{white!15!black},xshift=-5pt},
ylabel={Acceptance Ratio},
axis background/.style={fill=white},
xmajorgrids,
ymajorgrids,
yticklabel shift=-2pt,
ylabel shift=-5pt,
xlabel shift=-2pt,
ybar=0pt,
xtick align=inside
]
\addplot[ybar, bar width=0.25, fill=Reds-F, draw=Reds-F, area legend] table[row sep=crcr] {%
3	0.968\\
4	0.936\\
5	0.52\\
};
\addplot[forget plot, color=white!15!black] table[row sep=crcr] {%
2.51111111111111	0\\
5.48888888888889	0\\
};
\addplot[ybar, bar width=0.25, fill=Reds-I, draw=Reds-I, area legend] table[row sep=crcr] {%
3	1\\
4	1\\
5	1\\
};
\addplot[forget plot, color=white!15!black] table[row sep=crcr] {%
2.51111111111111	0\\
5.48888888888889	0\\
};
\addplot[ybar, bar width=0.25, fill=Reds-L, draw=Reds-L, area legend] table[row sep=crcr] {%
3	1\\
4	1\\
5	1\\
};
\addplot[forget plot, color=white!15!black] table[row sep=crcr] {%
2.51111111111111	0\\
5.48888888888889	0\\
};
\end{axis}

\begin{axis}[%
width=0.167\fwidth,
height=\fheight,
at={(0.253\fwidth,0\fheight)},
scale only axis,
bar shift auto,
xmin=2.51111111111111,
xmax=5.48888888888889,
xtick={3,4,5},
xticklabels={{300},{600},{1500}},
xticklabel style={rotate=90},
xlabel style={font=\footnotesize\color{white!15!black}, at={(1.265,-0.5)}},
xlabel={Number of Requested Instances $I$},
ymin=0,
ymax=1,
ylabel style={font=\footnotesize\color{white!15!black},xshift=-5pt},
ylabel={Activation Ratio},
axis background/.style={fill=white},
xmajorgrids,
ymajorgrids,
yticklabel shift=-2pt,
ylabel shift=-5pt,
xlabel shift=-2pt,
ybar=0pt,
xtick align=inside
]
\addplot[ybar, bar width=0.25, fill=Reds-F, draw=Reds-F, area legend] table[row sep=crcr] {%
3	0.854\\
4	0.926\\
5	0.882\\
};
\addplot[forget plot, color=white!15!black] table[row sep=crcr] {%
2.51111111111111	0\\
5.48888888888889	0\\
};
\addplot[ybar, bar width=0.25, fill=Reds-I, draw=Reds-I, area legend] table[row sep=crcr] {%
3	1\\
4	1\\
5	1\\
};
\addplot[forget plot, color=white!15!black] table[row sep=crcr] {%
2.51111111111111	0\\
5.48888888888889	0\\
};
\addplot[ybar, bar width=0.25, fill=Reds-L, draw=Reds-L, area legend] table[row sep=crcr] {%
3	0.1\\
4	0.1\\
5	0.1\\
};
\addplot[forget plot, color=white!15!black] table[row sep=crcr] {%
2.51111111111111	0\\
5.48888888888889	0\\
};
\end{axis}

\begin{axis}[%
width=0.167\fwidth,
height=\fheight,
at={(0.506\fwidth,0\fheight)},
scale only axis,
bar shift auto,
xmin=2.51111111111111,
xmax=5.48888888888889,
xtick={3,4,5},
xticklabels={{300},{600},{1500}},
xticklabel style={rotate=90},
xlabel style={font=\footnotesize\color{white!15!black}},
ymin=0,
ymax=30,
ylabel style={font=\footnotesize\color{white!15!black}},
ylabel={Energy [kJ]},
axis background/.style={fill=white},
xmajorgrids,
ymajorgrids,
yticklabel shift=-2pt,
ylabel shift=-5pt,
xlabel shift=-2pt,
ybar=0pt,
xtick align=inside
]
\addplot[ybar, bar width=0.25, fill=Reds-F, draw=Reds-F, area legend] table[row sep=crcr] {%
3	7.60173599999829\\
4	12.0889440000031\\
5	15.3353999999949\\
};
\addplot[forget plot, color=white!15!black] table[row sep=crcr] {%
2.51111111111111	0\\
5.48888888888889	0\\
};
\addplot[ybar, bar width=0.25, fill=Reds-I, draw=Reds-I, area legend] table[row sep=crcr] {%
3	8.27700000000804\\
4	12.9539999999973\\
5	26.985000000004\\
};
\addplot[forget plot, color=white!15!black] table[row sep=crcr] {%
2.51111111111111	0\\
5.48888888888889	0\\
};
\addplot[ybar, bar width=0.25, fill=Reds-L, draw=Reds-L, area legend] table[row sep=crcr] {%
3	5.03700000001056\\
4	9.71399999999986\\
5	23.7450000000065\\
};
\addplot[forget plot, color=white!15!black] table[row sep=crcr] {%
2.51111111111111	0\\
5.48888888888889	0\\
};
\end{axis}

\begin{axis}[%
width=0.167\fwidth,
height=\fheight,
at={(0.759\fwidth,0\fheight)},
scale only axis,
bar shift auto,
xmin=2.51111111111111,
xmax=5.48888888888889,
xtick={3,4,5},
xticklabels={{300},{600},{1500}},
xticklabel style={rotate=90},
xlabel style={font=\footnotesize\color{white!15!black}},
ymin=0,
ymax=80,
ylabel style={font=\footnotesize\color{white!15!black}},
ylabel={Latency [s]},
axis background/.style={fill=white},
xmajorgrids,
ymajorgrids,
yticklabel shift=-2pt,
ylabel shift=-5pt,
xlabel shift=-2pt,
legend columns=3,
legend style={legend cell align=left, align=left, draw=white!15!black,
anchor=south, font=\scriptsize, align=left, at={(-1.75,1.1)}},
ybar=0pt,
xtick align=inside
]
\addplot[ybar, bar width=0.25, fill=Reds-F, draw=Reds-F, area legend] table[row sep=crcr] {%
3	0.541274508\\
4	1.28777598266667\\
5	1.98607191466667\\
};
\addplot[forget plot, color=white!15!black] table[row sep=crcr] {%
2.51111111111111	0\\
5.48888888888889	0\\
};
\addlegendentry{\name}

\addplot[ybar, bar width=0.25, fill=Reds-I, draw=Reds-I, area legend] table[row sep=crcr] {%
3	0.537666666666667\\
4	1.34360666666667\\
5	6.14180666666667\\
};
\addplot[forget plot, color=white!15!black] table[row sep=crcr] {%
2.51111111111111	0\\
5.48888888888889	0\\
};
\addlegendentry{Load Balancing}

\addplot[ybar, bar width=0.25, fill=Reds-L, draw=Reds-L, area legend] table[row sep=crcr] {%
3	14.1388066666667\\
4	30.1328066666667\\
5	78.1148066666667\\
};
\addplot[forget plot, color=white!15!black] table[row sep=crcr] {%
2.51111111111111	0\\
5.48888888888889	0\\
};
\addlegendentry{No Scaling}

\draw[draw=black] (axis cs:2.51111111111111,0) rectangle (axis cs:5.48888888888889,8);

\addplot[color=black!40, smooth]
  table[row sep=crcr]{%
2.51111111111111 8\\
3.06 35\\
};

\addplot [color=black!40, smooth]
  table[row sep=crcr]{%
5.48888888888889 8\\
5.03 71.5\\
};

\end{axis}

\begin{axis}[%
width=0.11\fwidth,
height=0.45\fheight,
at={(0.79\fwidth,0.44\fheight)},
scale only axis,
bar shift auto,
xmin=2.51111111111111,
xmax=5.48888888888889,
xtick={3,4,5},
xticklabels={,,,},
xticklabel style={rotate=90},
xlabel style={font=\footnotesize\color{white!15!black}},
ymin=0,
ymax=8,
ytick={0,4,8},
ylabel style={font=\footnotesize\color{white!15!black}},
axis background/.style={fill=white},
xmajorgrids,
ymajorgrids,
yticklabel shift=-2pt,
ylabel shift=-5pt,
xlabel shift=-2pt,
legend columns=3,
legend style={legend cell align=left, align=left, draw=white!15!black,
anchor=south, font=\scriptsize, align=left, at={(-1.75,1.1)}},
ybar=0pt,
xtick align=inside
]
\addplot[ybar, bar width=0.25, fill=Reds-F, draw=Reds-F, area legend] table[row sep=crcr] {%
3   0.541274508\\
4   1.28777598266667\\
5   1.98607191466667\\
};
\addplot[forget plot, color=white!15!black] table[row sep=crcr] {%
2.51111111111111    0\\
5.48888888888889    0\\
};
\addlegendentry{\name}

\addplot[ybar, bar width=0.25, fill=Reds-I, draw=Reds-I, area legend] table[row sep=crcr] {%
3   0.537666666666667\\
4   1.34360666666667\\
5   6.14180666666667\\
};
\addplot[forget plot, color=white!15!black] table[row sep=crcr] {%
2.51111111111111    0\\
5.48888888888889    0\\
};
\addlegendentry{Load Balancing}

\addplot[ybar, bar width=0.25, fill=Reds-L, draw=Reds-L, area legend] table[row sep=crcr] {%
3   14.1388066666667\\
4   30.1328066666667\\
5   78.1148066666667\\
};
\addplot[forget plot, color=white!15!black] table[row sep=crcr] {%
2.51111111111111    0\\
5.48888888888889    0\\
};
\addlegendentry{No Scaling}

\legend{}  

\end{axis}

\end{tikzpicture}%
    \setlength\belowcaptionskip{-.3cm}
    \caption{\name vs.\ others for $S=10$ and varying number of instances~($I$).}
    \label{fig:comparison}
\end{figure}
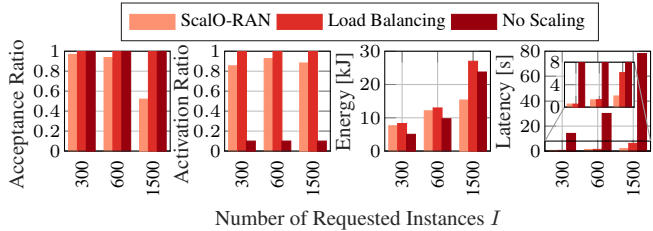

Finally, Fig.~\ref{fig:comparison} compare \name against two other approaches, i.e., resource-based load balancing (native in OpenShift and frequently considered in the literature~\cite{vaquero2011dynamically,bauer2019chamulteon,gulati2011cloud}) and no scaling, for $S=10$.
Load balancing distributes requests among servers based on congestion levels, while with no scaling all requests are instantiated on a single server. Load balancing and no scaling always admit all requests, while \name accepts $\sim$98\% of requests when $I=300$ and 52\% when $I=1500$. Moreover, no scaling activates one server only, load balancing activates all servers, and \name activates on average 90\% of servers. The lower \name acceptance and activation ratios are not a drawback, but a consequence of the energy-aware profit maximization objective coupled with the maximum inference time requirement. Together, these force \name to accept and distribute only those requests that guarantee timely inference time as requested by tenants.

Indeed, we see that no scaling is not suitable for \oran applications due to the extremely high inference time (Fig.~\ref{fig:comparison}). If compared to load balancing, \name provides a lower energy consumption and a lower inference time, which also satisfies timing requirements from tenants. Overall, \name performs better than widely used load balancing approaches by reducing energy while guaranteeing a timely inference.

\ifexttikz
    \tikzsetnextfilename{coexistence}
\fi


\ifexttikz
    \tikzsetnextfilename{comparison}
\fi


\section{Experimental Results}
\label{sec:results}%

We use the prototype of Sec.~\ref{section:framework} to experimentally evaluate \name and compare it against load balancing policies of OpenShift. 
We consider the same setup from Sec.~\ref{section:numerical-evaluation} with the three \ai-based xApps in Fig.~\ref{fig:scatter-plot-single-xapps}, the average fit in Table~\ref{table:piecewise} and $I=123$ xApp instances to be deployed. Our prototype embeds two Dell PowerEdge R340 worker nodes (i.e., $\mathrm{WN}_1$, $\mathrm{WN}_2$, Fig.~\ref{figure:prototype}) used for xApp deployment,
thus we consider $R=3$ tenants with one request each ($r_1$, $r_2$ and $r_3$) to mimic a small \oran deployment.
We consider \nearrt inference time profiles ($[350,1000]$\:ms), and only $r_1$ demands the maximum inference time of $350$\:ms. The monetary value is $\rho_{r_1} = 30 \rho_{r_2} = 30 \rho_{r_3}$.
%

\ifexttikz
    \tikzsetnextfilename{exp-resources}
\fi 
\begin{figure}[t]
    \setlength\abovecaptionskip{0pt}
    \centering
    \setlength\fwidth{0.95\columnwidth}
    \setlength\fheight{0.2\columnwidth}
    \includegraphics[width=0.95\columnwidth]{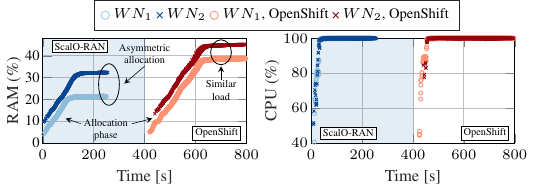}
    \caption{RAM and CPU utilization over time with \name and OpenShift.}
    \label{fig:exp-resources}
\vspace{-5pt}
\end{figure}

Fig.~\ref{fig:exp-resources} shows the CPU and RAM utilization over time for both \name and OpenShift for a single 4-minute experiment. Here, \name admits only requests $r_1$ and $r_2$ (demanding $350$\:ms and $1000$\:ms, respectively) instantiating $82$ xApps, while OpenShift admits all requests and all $123$ xApps. We notice that OpenShift allocates all xApps evenly across $\mathrm{WN}_1$ and $\mathrm{WN}_2$ due to load balancing. Instead, \name allocates instances in a more asymmetric way. Specifically, 85\% of xApps on $\mathrm{WN}_1$ are from $r_1$, and the remaining 15\% is from $r_2$. Instead, 100\% of xApps on $\mathrm{WN}_2$ are from $r_2$. This allocation, especially the allocation on $\mathrm{WN}_1$, ensures that all xApps satisfy the $350$\:ms inference constraint on $\mathrm{WN}_1$ as required by $r_1$. Instead, CPU usage is almost 100\% except for the initial deployment phase. The allocation phase is voluntarily slow, as we allocate one xApp at a time to collect reliable data.

\ifexttikz
    \tikzsetnextfilename{exp-inference-time}
\fi 
\begin{figure}[t]
    \setlength\abovecaptionskip{0pt}
    \centering
    \setlength\fwidth{0.95\columnwidth}
    \setlength\fheight{0.2\columnwidth}
    \includegraphics[width=0.95\columnwidth]{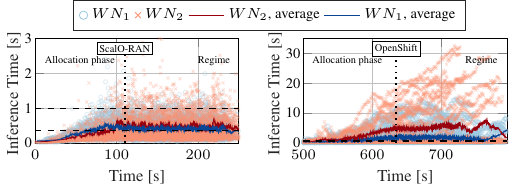}
        \setlength\belowcaptionskip{-.3cm}
    \caption{Evolution of the inference time with \name and OpenShift.}
    \label{fig:exp-inference-time}
\end{figure}

Fig.~\ref{fig:exp-inference-time} reports the inference time over time for \name and OpenShift. We clearly see that OpenShift cannot satisfy even the $1000$\:ms requirement, as it allocates all xApps without considering their timing requirements. This results in inference time violations that affect the proper functioning of the \ran. Instead, \name not only admits requests whose demands can be accommodated, but distributes xApps to ensure that $\mathrm{WN}_1$ (which hosts all xApps of $r_1$ and 15\% of xApps of $r_2$) delivers the $350$\:ms requirements on average, while $\mathrm{WN}_2$ can guarantee the $1000$\:ms requirement from $r_2$.
\ifexttikz
    \tikzsetnextfilename{exp-stats}
\fi 
\begin{figure}[h]
    \setlength\abovecaptionskip{0pt}
    \centering
    \setlength\fwidth{0.95\columnwidth}
    \setlength\fheight{0.35\columnwidth}
    \input{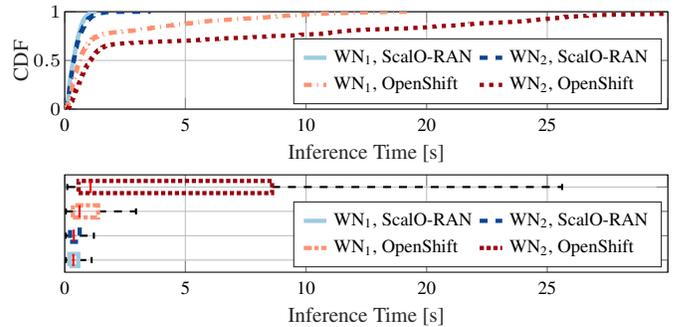}
    \caption{CDF and boxplot of inference time with \name and OpenShift.}
    \label{fig:exp-stats}
\end{figure}
Finally, in Fig.~\ref{fig:exp-stats} we show the \gls{cdf} for the different worker nodes and approaches, as well as the boxplots showing median values. We see that OpenShift cannot guarantee any inference time demand, while \name ensures that the expected inference time follows tenant requirements.


\section{Conclusions}
\label{section:conclusions}


In this work, we presented \name, an O-RAN energy-aware scaling system to enforce inference time constraints on intelligent applications. We provided a latency model based on a measurement campaign on an OpenShift cluster, a mathematical optimization model, and an O-RAN-compliant prototype. We compared \name with OpenShift's scaling mechanism, showing that \name is able to deploy O-RAN applications complying with the operators' latency constraints.
%
Results demonstrate that scaling \ai solutions in \oran is not resource-constrained only, but also time-constrained as requirements on the inference time strongly affect how many dApps, xApps and rApps can coexist on the same server.


\balance
\footnotesize
\bibliographystyle{IEEEtran}
\bibliography{bibliography.bib} 

\begin{thebibliography}{10}
\providecommand{\url}[1]{#1}
\csname url@samestyle\endcsname
\providecommand{\newblock}{\relax}
\providecommand{\bibinfo}[2]{#2}
\providecommand{\BIBentrySTDinterwordspacing}{\spaceskip=0pt\relax}
\providecommand{\BIBentryALTinterwordstretchfactor}{4}
\providecommand{\BIBentryALTinterwordspacing}{\spaceskip=\fontdimen2\font plus
\BIBentryALTinterwordstretchfactor\fontdimen3\font minus
  \fontdimen4\font\relax}
\providecommand{\BIBforeignlanguage}[2]{{%
\expandafter\ifx\csname l@#1\endcsname\relax
\typeout{** WARNING: IEEEtran.bst: No hyphenation pattern has been}%
\typeout{** loaded for the language `#1'. Using the pattern for}%
\typeout{** the default language instead.}%
\else
\language=\csname l@#1\endcsname
\fi
#2}}
\providecommand{\BIBdecl}{\relax}
\BIBdecl

\bibitem{polese2022understanding}
M.~Polese, L.~Bonati, S.~D’Oro, S.~Basagni, and T.~Melodia, ``{Understanding
  {O-RAN}: Architecture, Interfaces, Algorithms, Security, and Research
  Challenges},'' \emph{IEEE Communications Surveys \& Tutorials}, 2023.

\bibitem{brik2022deep}
B.~Brik, K.~Boutiba, and A.~Ksentini, ``{Deep Learning for B5G Open Radio
  Access Network: Evolution, Survey, Case Studies, and Challenges},''
  \emph{IEEE Open Journal of the Communications Society}, vol.~3, 2022.

\bibitem{doro2022dapps}
S.~D'Oro, M.~Polese, L.~Bonati, H.~Cheng, and T.~Melodia, ``{dApps: Distributed
  Applications for Real-Time Inference and Control in O-RAN},'' \emph{IEEE
  Communications Magazine}, vol.~60, no.~11, 2022.

\bibitem{schmidt2021ran}
R.~Schmidt and N.~Nikaein, ``{RAN Engine: Service-Oriented RAN Through
  Containerized Micro-Services},'' \emph{IEEE Transactions on Network and
  Service Management}, vol.~18, no.~1, pp. 469--481, March 2021.

\bibitem{niknam2022intelligent}
S.~Niknam, A.~Roy, H.~S. Dhillon, S.~Singh, R.~Banerji, J.~H. Reed, N.~Saxena,
  and S.~Yoon, ``{Intelligent O-RAN for Beyond 5G and 6G Wireless Networks},''
  in \emph{IEEE Globecom Workshops}, December 2022.

\bibitem{bonati2021intelligence}
L.~Bonati, S.~D'Oro, M.~Polese, S.~Basagni, and T.~Melodia, ``{Intelligence and
  Learning in O-RAN for Data-driven NextG Cellular Networks},'' \emph{IEEE
  Communications Magazine}, vol.~59, no.~10, pp. 21--27, October 2021.

\bibitem{ojiaghi2023benefits}
B.~Ojaghi, F.~Adelantado, and C.~Verikoukis, ``{On the Benefits of vDU
  Standardization in Softwarized NG-RAN: Enabling Technologies, Challenges, and
  Opportunities},'' \emph{IEEE Communications Magazine}, vol.~61, no.~4, pp.
  92--98, April 2023.

\bibitem{garciaaviles2021nuberu}
G.~Garcia-Aviles, A.~Garcia-Saavedra, M.~Gramaglia, X.~Costa-Perez, P.~Serrano,
  and A.~Banchs, ``{Nuberu: Reliable {RAN} Virtualization in Shared
  Platforms},'' in \emph{Proc. of ACM MobiCom}, October 2021.

\bibitem{garcia2018fluidran}
A.~Garcia-Saavedra, X.~Costa-Perez, D.~J. Leith, and G.~Iosifidis, ``{FluidRAN:
  Optimized vRAN/MEC Orchestration},'' in \emph{Proc. of IEEE INFOCOM}, April
  2018.

\bibitem{parvez2018survey}
I.~Parvez, A.~Rahmati, I.~Guvenc, A.~I. Sarwat, and H.~Dai, ``{A Survey on Low
  Latency Towards 5G: RAN, Core Network and Caching Solutions},'' \emph{IEEE
  Communications Surveys \& Tutorials}, vol.~20, no.~4, 2018.

\bibitem{giannone2019impact}
F.~Giannone, K.~Kondepu, H.~Gupta, F.~Civerchia, P.~Castoldi,
  A.~Antony~Franklin, and L.~Valcarenghi, ``{Impact of Virtualization
  Technologies on Virtualized RAN Midhaul Latency Budget: A Quantitative
  Experimental Evaluation},'' \emph{IEEE Communications Letters}, vol.~23,
  no.~4, pp. 604--607, April 2019.

\bibitem{bonati2023neutran}
L.~Bonati, M.~Polese, S.~D'Oro, S.~Basagni, and T.~Melodia, ``{NeutRAN: An Open
  RAN Neutral Host Architecture for Zero-Touch RAN and Spectrum Sharing},''
  \emph{IEEE Transactions on Mobile Computing}, 2023.

\bibitem{polese2021machine}
M.~Polese, R.~Jana, V.~Kounev, K.~Zhang, S.~Deb, and M.~Zorzi, ``{Machine
  Learning at the Edge: A Data-Driven Architecture With Applications to 5G
  Cellular Networks},'' \emph{IEEE Transactions on Mobile Computing}, vol.~20,
  no.~12, pp. 3367--3382, December 2021.

\bibitem{vaquero2011dynamically}
L.~M. Vaquero, L.~Rodero-Merino, and R.~Buyya, ``{Dynamically Scaling
  Applications in the Cloud},'' \emph{SIGCOMM Compututer Communication Review},
  vol.~41, no.~1, p. 45–52, January 2011.

\bibitem{bauer2019chamulteon}
A.~Bauer, V.~Lesch, L.~Versluis, A.~Ilyushkin, N.~Herbst, and S.~Kounev,
  ``{Chamulteon: Coordinated Auto-Scaling of Micro-Services},'' in \emph{Proc.
  of IEEE ICDCS}, July 2019.

\bibitem{gulati2011cloud}
A.~Gulati, G.~Shanmuganathan, A.~Holler, and I.~Ahmad, ``{Cloud Scale Resource
  Management: Challenges and Techniques},'' in \emph{Proc. of USENIX HotCloud},
  2011.

\bibitem{singhvi2021atoll}
A.~Singhvi, A.~Balasubramanian, K.~Houck, M.~D. Shaikh, S.~Venkataraman, and
  A.~Akella, ``{Atoll: A Scalable Low-Latency Serverless Platform},'' in
  \emph{Proc. of ACM Symposium on Cloud Computing}, 2021.

\bibitem{hu2021k8s}
T.~Hu and Y.~Wang, ``{A Kubernetes Autoscaler Based on Pod Replicas
  Prediction},'' in \emph{Asia-Pacific Conference on Communications Technology
  and Computer Science (ACCTCS)}, Jan 2021, pp. 238--241.

\bibitem{rossi2019horizontal}
F.~Rossi, M.~Nardelli, and V.~Cardellini, ``{Horizontal and Vertical Scaling of
  Container-Based Applications Using Reinforcement Learning},'' in \emph{Proc.
  of IEEE CLOUD}, July 2019.

\bibitem{casalicchio2017auto}
E.~Casalicchio and V.~Perciballi, ``{Auto-Scaling of Containers: The Impact of
  Relative and Absolute Metrics},'' in \emph{Proc. of IEEE FAS*W}, 2017.

\bibitem{ABENI2020101709}
L.~Abeni and D.~Faggioli, ``{Using Xen and KVM as Real-time Hypervisors},''
  \emph{Journal of Systems Architecture}, vol. 106, p. 101709, 2020.

\bibitem{distefano2020ananke}
A.~Di~Stefano, A.~Di~Stefano, and G.~Morana, ``{Ananke: A Framework for
  Cloud-Native Applications Smart Orchestration},'' in \emph{Proc. of IEEE
  WETICE}, 2020.

\bibitem{prachitmutita}
I.~Prachitmutita, W.~Aittinonmongkol, N.~Pojjanasuksakul, M.~Supattatham, and
  P.~Padungweang, ``{Auto-scaling Microservices on IaaS Under SLA with
  Cost-effective Framework},'' in \emph{Proc. of IEEE ICACI}, 2018.

\bibitem{han2012lightweigth}
R.~Han, L.~Guo, M.~M. Ghanem, and Y.~Guo, ``{Lightweight Resource Scaling for
  Cloud Applications},'' in \emph{Proc. of IEEE/ACM CCGrid}, 2012.

\bibitem{k8sscaling}
\BIBentryALTinterwordspacing
{Kubernetes}, ``{Horizontal Pod Autoscaling},'' 2023. [Online]. Available:
  \url{https://tinyurl.com/5n8jykm3}
\BIBentrySTDinterwordspacing

\bibitem{mao2010cloud}
M.~Mao, J.~Li, and M.~Humphrey, ``{Cloud Auto-scaling with Deadline and Budget
  Constraints},'' in \emph{Proc. of IEEE/ACM International Conference on Grid
  Computing}, 2010.

\bibitem{mao2011auto}
M.~Mao and M.~Humphrey, ``{Auto-Scaling to Minimize Cost and Meet Application
  Deadlines in Cloud Workflows},'' in \emph{Proc. of International Conference
  for High Performance Computing, Networking, Storage and Analysis}, 2011.

\bibitem{anagnostou2019towards}
A.~Anagnostou, S.~J.~E. Taylor, N.~Tijjani~Abubakar, T.~Kiss, J.~DesLauriers,
  G.~Gesmier, G.~Terstyanszky, P.~Kacsuk, and J.~Kovacs, ``{Towards a
  Deadline-Based Simulation Experimentation Framework Using Micro-Services
  Auto-Scaling Approach},'' in \emph{Proc. of IEEE WSC}, 2019.

\bibitem{das2016automated}
S.~Das, F.~Li, V.~R. Narasayya, and A.~C. K\"{o}nig, ``Automated demand-driven
  resource scaling in relational database-as-a-service,'' in \emph{Proc. ACM
  International Conference on Management of Data}, 2016.

\bibitem{dryjanskiran}
\BIBentryALTinterwordspacing
M.~Hoffmann and M.~Dryja{\'n}ski, ``{The O-RAN Whitepaper 2023: Energy
  Efficiency in O-RAN},'' Rimedo Labs White Paper. [Online]. Available:
  \url{https://tinyurl.com/ys7mmk69}
\BIBentrySTDinterwordspacing

\bibitem{fei2018adaptive}
X.~Fei, F.~Liu, H.~Xu, and H.~Jin, ``{Adaptive VNF Scaling and Flow Routing
  with Proactive Demand Prediction},'' in \emph{Proc. of IEEE INFOCOM}, April
  2018.

\bibitem{8761272}
Y.~Bi, C.~Colman-Meixner, R.~Wang, F.~Meng, R.~Nejabati, and D.~Simeonidou,
  ``{Resource Allocation for Ultra-Low Latency Virtual Network Services in
  Hierarchical 5G Network},'' in \emph{Proc. of IEEE ICC}, 2019.

\bibitem{9463941}
D.~Harutyunyan, R.~Behravesh, and N.~Slamnik-Kriještorac, ``{Cost-efficient
  Placement and Scaling of 5G Core Network and MEC-enabled Application VNFs},''
  in \emph{Proc. of IFIP/IEEE International Symposium on Integrated Network
  Management (IM)}, 2021.

\bibitem{ali2023proactive}
K.~Ali and M.~Jammal, ``{Proactive VNF Scaling and Placement in 5G O-RAN using
  ML},'' \emph{IEEE Transactions on Network and Service Management}, 2023.

\bibitem{doro2022orchestran}
S.~D'Oro, L.~Bonati, M.~Polese, and T.~Melodia, ``{OrchestRAN: Network
  Automation through Orchestrated Intelligence in the Open RAN},'' in
  \emph{{Proc. of IEEE INFOCOM}}, May 2022.

\bibitem{thaliath2022predictive}
J.~Thaliath, S.~Niknam, S.~Singh, R.~Banerji, N.~Saxena, H.~S. Dhillon, J.~H.
  Reed, A.~K. Bashir, A.~Bhat, and A.~Roy, ``{Predictive Closed-Loop Service
  Automation in O-RAN Based Network Slicing},'' \emph{IEEE Communications
  Standards Magazine}, vol.~6, no.~3, pp. 8--14, Sep. 2022.

\bibitem{7864818}
A.~Capone, S.~D'Elia, I.~Filippini, A.~E.~C. Redondi, and M.~Zangani,
  ``Modeling energy consumption of mobile radio networks: An operator
  perspective,'' \emph{IEEE Wireless Communications}, vol.~24, no.~4, 2017.

\bibitem{experimental-pw-vbs}
J.~A. Ayala-Romero, I.~Khalid, A.~Garcia-Saavedra, X.~Costa-Perez, and
  G.~Iosifidis, ``{Experimental Evaluation of Power Consumption in Virtualized
  Base Stations},'' in \emph{Proc. of IEEE ICC}, 2021.

\bibitem{joint-opt-energy-latency}
T.~Pamuklu, S.~Mollahasani, and M.~Erol-Kantarci, ``{Energy-Efficient and
  Delay-Guaranteed Joint Resource Allocation and {DU} Selection in O-{RAN}},''
  in \emph{Proc. of IEEE 5GWF}, October 2021.

\bibitem{bonati2020cellos}
L.~Bonati, S.~D'Oro, L.~Bertizzolo, E.~Demirors, Z.~Guan, S.~Basagni, and
  T.~Melodia, ``{CellOS: Zero-touch Softwarized Open Cellular Networks},''
  \emph{Computer Networks}, vol. 180, pp. 1--13, October 2020.

\bibitem{near-rt-ric-installation}
``{Installation of the OSC Near-real-time RIC},''
  \url{https://shorturl.at/sv135}.

\bibitem{polese2022coloran}
M.~Polese, L.~Bonati, S.~D'Oro, S.~Basagni, and T.~Melodia, ``{ColO-RAN:
  Developing Machine Learning-based xApps for Open RAN Closed-loop Control on
  Programmable Experimental Platforms},'' \emph{IEEE Transactions on Mobile
  Computing}, pp. 1--14, July 2022.

\bibitem{dunham1986optimum}
J.~G. Dunham, ``{Optimum Uniform Piecewise Linear Approximation of Planar
  Curves},'' \emph{IEEE Transactions on Pattern Analysis and Machine
  Intelligence}, no.~1, pp. 67--75, 1986.

\bibitem{vieth1989fitting}
E.~Vieth, ``Fitting piecewise linear regression functions to biological
  responses,'' \emph{Journal of applied physiology}, vol.~67, no.~1, 1989.

\bibitem{ngueveu2019piecewise}
S.~U. Ngueveu, ``{Piecewise Linear Bounding of Univariate Nonlinear Functions
  and Resulting Mixed Integer Linear Programming-based Solution Methods},''
  \emph{European Journal of Operational Research}, vol. 275, no.~3, pp.
  1058--1071, 2019.

\bibitem{fan2007power}
X.~Fan, W.-D. Weber, and L.~A. Barroso, ``{Power Provisioning for a
  Warehouse-sized Computer},'' \emph{ACM SIGARCH Computer Architecture News},
  vol.~35, no.~2, pp. 13--23, 2007.

\bibitem{klimm2022packing}
M.~Klimm, M.~E. Pfetsch, R.~Raber, and M.~Skutella, ``Packing under convex
  quadratic constraints,'' \emph{Mathematical Programming}, vol. 192, no. 1-2,
  pp. 361--386, 2022.

\bibitem{baskin2018streaming}
C.~Baskin, N.~Liss, E.~Zheltonozhskii, A.~M. Bronstein, and A.~Mendelson,
  ``{Streaming Architecture for Large-scale Quantized Neural Networks on an
  FPGA-based Dataflow Platform},'' in \emph{Proc. of IEEE IPDPSW}, 2018.

\end{thebibliography}
\end{document}